\documentclass[10pt]{article}
\pdfoutput=1
\usepackage[utf8]{inputenc}
\usepackage[T1]{fontenc}
\usepackage{textcomp}
\usepackage{amsmath}
\usepackage{txfonts}
\usepackage[letterpaper,margin=1in]{geometry}
\usepackage{graphicx}
\usepackage{booktabs}
\usepackage{array}
\usepackage{longtable}
\usepackage[font=small,labelfont=bf,labelsep=period,skip=6pt]{caption}
\usepackage{titlesec}
\usepackage{microtype}
\usepackage{xurl}
\usepackage[hidelinks]{hyperref}
\usepackage{fancyhdr}

\titleformat{\section}{\large\bfseries}{\thesection.}{0.5em}{}
\titleformat{\subsection}{\normalsize\bfseries}{\thesubsection}{0.6em}{}
\titlespacing*{\section}{0pt}{1.4em}{0.5em}
\titlespacing*{\subsection}{0pt}{1.1em}{0.3em}
\fancypagestyle{firstpage}{\fancyhf{}}
\hypersetup{
  pdftitle={Forecast Lead Time and the Timing of Crew Deployment for Weather-Driven Power Outages Using AI Weather Models},
  pdfauthor={Mohammad Kian Golkar, Luciano Alves de Oliveira, Mohamad Alipour},
  pdfkeywords={power outages, AI weather forecasting, crew deployment, outage prediction, conformal prediction, data leakage}
}

\begin{document}
\thispagestyle{firstpage}

\begin{center}
\vspace*{0.2in}
\rule{\linewidth}{0.6pt}\par\vspace{0.12in}
{\Large\bfseries\scshape Forecast Lead Time and the Timing of Crew Deployment for\\[2pt] Weather-Driven Power Outages Using AI Weather Models\par}
\vspace{0.1in}\rule{\linewidth}{0.6pt}\par
\vspace{0.3in}
{\large\bfseries Mohammad Kian Golkar\textsuperscript{1}, Luciano Alves de Oliveira\textsuperscript{1,*}, Mohamad Alipour\textsuperscript{2}\par}
\vspace{0.12in}
{\small
\textsuperscript{1}Department of Agricultural and Biological Engineering, University of Illinois Urbana-Champaign, Urbana, IL 61801, USA\par
\vspace{0.04in}
\textsuperscript{2}Department of Civil Engineering, The Grainger College of Engineering, University of Illinois Urbana-Champaign, Urbana, IL 61821, USA\par
\vspace{0.1in}
\textsuperscript{*}Corresponding Author: \href{mailto:Luciano8@illinois.edu}{Luciano8@illinois.edu}\par}
\end{center}

\vspace{0.15in}
\begin{center}{\large\bfseries\scshape Abstract}\end{center}
\vspace{-0.4em}
\begin{quote}
\small
Utilities request mutual aid crews to restore power after severe storms, and requesting early secures more crews but relies on less accurate forecasts. We examined how forecast lead time affects the value of outage predictions for crew deployment, using operational forecasts from two AI weather models, GraphCast and FourCastNet, to predict outage occurrence for counties across the contiguous United States 24 to 120 hours ahead. Forecasts were evaluated by the share of outage burden captured by the counties where crews would be staged. Captured burden declined with lead time: the loss between three and five days was significant in all analyses, whereas the loss within three days was significant only for outages that had not yet started. Early deployment with all requested crews outperformed late deployment with fewer, and we report the availability ratio at which the two break even. Antecedent land surface predictors computed at forecast valid time include weather from the forecast period, which inflates their skill by 42 to 53\% and reverses their apparent advantage over forecast meteorology at long lead times. Computed correctly, they improved occurrence skill by 17.5\% at 24 hours but did not significantly change captured burden, a gap that attribution analysis linked to their static component. Prediction intervals covered only 0.476 of tropical cyclone events against a nominal 0.90. Calibration by hazard class corrected most of this shortfall at an 11\% cost in interval width, as a diagnostic only, because hazard class is known only after the event.
\end{quote}

\noindent\textbf{Keywords}\quad Power outages \textperiodcentered{} AI weather forecasting \textperiodcentered{} Crew deployment \textperiodcentered{} Outage prediction \textperiodcentered{} Conformal prediction \textperiodcentered{} Data leakage

\section{Introduction}

Power outages caused by severe weather lead to large economic and social losses, and their duration, frequency and magnitude have increased in recent years (Ma et al., 2025). Outages linked to climate hazards are widespread across the United States and affect socially vulnerable communities more than others (Do et al., 2023; Flores et al., 2024; Ganz et al., 2023). Their consequences include disrupted medical care and exposure to extreme heat and cold (Andresen et al., 2023), which makes the resilience of power systems under climate change an important concern (Panteli and Mancarella, 2015; Ptak et al., 2025). Recent studies have also shown that hardening the most critical circuits can improve this resilience (Stürmer et al., 2024).

Before a forecast storm, utilities position restoration crews in the areas expected to be affected (Guikema et al., 2014). For large events, they also request additional crews and restoration material from other utilities through mutual assistance agreements (Zhai et al., 2021). This is a costly decision, which can reach tens of millions of dollars per day for large utilities (Zhai et al., 2021). Crews must travel from other regions and be staged at specific locations in advance, so the decision fixes both the number of crews and their locations. Previous studies have optimised the dispatch of repair crews after an event together with network reconfiguration and mobile generation (Lei et al., 2019), and the staging of resources before hurricanes under uncertain scenarios (Arab et al., 2015). However, these studies take the forecast as given. The question of when the forecast is accurate enough to commit crews has not been studied as a decision in its own right. Requesting crews early secures them before other utilities do, but the allocation is based on a forecast several days ahead. Requesting late provides a better forecast, but the available crews may already be committed elsewhere. Both effects can be measured. Forecast quality at each lead time can be compared with observed outages, and the value of an allocation can be measured as the share of observed outage burden in the counties where crews were placed. Identifying the lead time at which forecast errors start to outweigh the benefit of early commitment would give utilities a quantitative basis for a decision that is currently based on convention.

Many statistical and machine learning models have been developed to predict weather-related outages. Most combine numerical weather prediction output with information on vegetation, infrastructure and the local environment. For distribution networks, these include decision tree and ensemble methods (Wanik et al., 2015; He et al., 2017; Cerrai et al., 2019; Watson et al., 2022), quantification of uncertainty in model training (Yang et al., 2020), hybrid models that combine physics and data for infrastructure hardening (Hughes et al., 2022), and, more recently, graph neural networks and methods for rare events (Shen et al., 2026; Azizi et al., 2026). For hurricanes, generalised additive and tree models have been used to estimate damage before landfall (Han et al., 2009; Guikema et al., 2010), to handle the large number of areas with zero outages (Guikema and Quiring, 2012), and to include soil, elevation and other local factors (McRoberts et al., 2018). Nateghi et al. (2014) showed that simpler models often perform as well as more complex ones. Similar models have been developed for extratropical storms in Europe (Tervo et al., 2021), and synthetic distribution systems have been used to predict outages where outage records are not available (Zhai et al., 2021).

These models have been useful in practice, but they share two limitations. First, they are trained on outage records from one utility or a small group of utilities, so they cover a service territory rather than a whole country. Second, their weather inputs come from conventional numerical weather prediction, usually the Weather Research and Forecasting model, or from reanalysis. The public release of the Environment for Analysis of Geo-Located Energy Information (EAGLE-I) outage record (Brelsford et al., 2024; Tansakul et al., 2023) has addressed the first limitation. This record has been used for national analysis of outages (Do et al., 2023), vulnerability indexing (Ma et al., 2025), prediction at state and regional scale (Fatehi et al., 2025), and data infrastructure that links outages to other geospatial data (Frakes et al., 2026). The second limitation remains, although weather forecasting itself has changed considerably in recent years.

Weather forecasting has changed rapidly in recent years with the development of AI models that are trained on reanalysis data instead of solving the equations of atmospheric motion. GraphCast produces forecasts up to ten days ahead that are competitive with operational systems for many variables (Lam et al., 2023). Pangu-Weather achieves similar results with a 3D transformer (Bi et al., 2023), and FourCastNet uses Fourier neural operators to forecast at high resolution (Pathak et al., 2022). Probabilistic models based on diffusion (Price et al., 2025) and general Earth system models (Bodnar et al., 2025) have followed, and archives of their forecasts are now available for community evaluation (Radford et al., 2025). However, the use of these forecasts for impact prediction has received much less attention than their meteorological evaluation. This matters because AI forecasts differ from conventional ones in ways that affect impact models. They smooth weather fields and weaken extremes as lead time increases, and they show systematic biases in tropical cyclone intensity (DeMaria et al., 2025). Charlton-Perez et al. (2024) found that they capture the overall structure of storms but underestimate peak wind speeds. Both properties affect damage prediction, because damage increases sharply with wind speed.

Several studies have examined how weather forecast errors propagate into outage predictions, but only for conventional forecasts. Yang et al. (2021) quantified how forecast uncertainty at lead times of one to five days affects outage prediction error. They found that the error depended weakly on lead time up to three days and increased stepwise at four and five days. Esmaeilian et al. (2026) examined error propagation under forecast, analysis and operational scenarios. These studies show that the question can be answered for conventional forecasts at the scale of a single utility. In this study, we address it for AI forecasts at national scale and extend it to an operational decision.

Other studies have shown that outage risk at a given wind speed also depends on the state of the land surface before the storm. Saturated soil weakens root anchorage and makes trees easier to uproot, trees in leaf catch more wind than bare trees, and snow adds load to power lines and branches. Manning et al. (2025) found for windstorms in the United Kingdom that high antecedent rainfall and summer foliage increase outage probability at a given wind speed, and that ignoring these factors can underestimate the probability of large outages by a factor of two to five. In North America, soil and elevation characteristics (Quiring et al., 2011; McRoberts et al., 2018) and vegetation (Cerrai et al., 2019) have also improved outage models. These land surface predictors are attractive for forecasting because they describe conditions that already exist when the forecast is issued. They may therefore remain informative as weather forecasts become less accurate at longer lead times, and this expectation can be tested.

However, these predictors also carry a methodological risk. Antecedent conditions are usually computed over a rolling window. If this window ends at the time the prediction is valid for, rather than the time the forecast is issued, it includes weather that occurs during the forecast period. This happens whenever reanalysis data are joined to the outage target on a shared time index, and it produces no error or warning. The resulting predictor then partly observes the storm being predicted instead of describing conditions before it, and this contamination increases with lead time. This is a form of data leakage, which Kapoor and Narayanan (2023) showed is common in scientific studies that use machine learning and inflates their reported performance. To our knowledge, it has not been quantified for antecedent predictors in impact forecasting.

This study links an operational archive of AI weather forecasts to a national outage record and evaluates the forecasts by the decisions they support rather than by prediction accuracy alone. We model outage occurrence and severity for counties across the contiguous United States from 2022 to 2024, using forecasts from the NOAA/CIRA archive of AI weather predictions (Radford et al., 2025) at lead times of 24 to 120 hours. We then assess what these forecasts are worth to a utility that must decide when to request mutual assistance crews.

The study makes four contributions. First, it analyses when crews should be committed. For each lead time, we measure the share of observed outage burden captured by staging crews in the counties ranked highest by the model. Paired resampling shows that the loss between three and five days is significant in all analyses, while the loss within the first three days is significant only when burden is limited to outages that have not yet started. We also report the crew availability ratio at which early commitment stops paying off, which requires no assumption about how availability declines when the request is delayed. This extends the forecast timing framework that Regnier and Harr (2006) and Regnier (2008) developed for hurricane evacuation to the commitment of infrastructure resources. The two problems differ, because a delayed evacuation faces a growing hazard, whereas a delayed crew request faces a shrinking pool of crews. We also express this analysis in economic terms without assuming a value of lost load, using potential economic value across a range of cost--loss ratios (Murphy, 1977; Richardson, 2000, 2001; Wilks, 2001) and the break-even staging cost per county-day. Second, it identifies and quantifies temporal leakage in antecedent land surface predictors. Evaluating these predictors at the forecast valid time instead of the issue time inflates their skill substantially. The inflation is concentrated in rolling window quantities, whose values at valid time and at issue time become almost uncorrelated at five days. Correcting the leakage reverses the conclusion that land surface state outperforms forecast meteorology at long lead times. The diagnostic is simple and can be applied to any impact forecasting study that uses antecedent conditions.

Third, it shows that gains in forecast skill do not necessarily produce gains in decisions. Adding land surface predictors improves occurrence skill considerably but does not significantly change captured outage burden at any lead time. Attribution analysis (Lundberg and Lee, 2017) links this gap to the static component of the land surface group, whose share of attribution increases as the forecast degrades. The difference between forecast skill and forecast value is well known in forecast verification (Murphy, 1977; Richardson, 2000), but it has rarely been examined in outage prediction. Fourth, it shows that prediction intervals can fail when the mix of hazards changes. Intervals built by conformalized quantile regression (Romano et al., 2019) have correct overall coverage but fail badly for individual hazard classes, because the calibration year had no major hurricane landfall on the US mainland and the test year had three. Calibration by hazard class (Vovk, 2013) restores coverage at a small cost in interval width, but only as a diagnostic, since hazard class is assigned after the event. Adding the training year to the calibration sample reduces coverage more than the shift it was meant to correct.

The remainder of this paper is organised as follows. Section 2 describes the data, the predictive model, the evaluation methods and the decision analysis. Section 3 presents and discusses the results and the limitations of the study, and Section 4 concludes.

\section{Materials and methods}

\subsection*{Nomenclature}
{\small
\begin{longtable}{@{}p{0.18\linewidth}p{0.76\linewidth}@{}}
\toprule
\textit{Symbol} & \textit{Description} \\
\midrule
\endhead
\bottomrule
\endlastfoot
\(C\) & County index, \(c = 1,\ldots,C\) with \(C = 2491\) \\
\(t\) & Valid time of a 6-hourly window \\
\(L\) & Forecast lead time, \(L \in \{24, 48, 72, 120\}\)~h \\
\(n_{c,t}\) & Customers without power in county \(c\) at time \(t\) \\
\(H_{c}\) & Housing units in county \(c\) \\
\(s_{c,t}\) & Outage severity, \(n_{c,t}/H_{c}\) \\
\(y_{c,t}\) & Binary outage event indicator \\
\(\tau\) & Severity threshold defining an event \\
\(\Delta\) & Minimum event duration \\
\(g\) & Gap tolerance within an event \\
\(m_{c,t}^{(L)}\) & Meteorological feature vector, forecast at lead \(L\) \\
\(l_{c,t}\) & Land-surface feature vector at valid time \\
\(x_{c,t}^{(L)}\) & Complete predictor vector at lead \(L\) \\
\(\pi_{c,t}\) & Predicted occurrence probability \\
\(z_{t}\) & Data-regime indicator \\
\(\mu_{c,t}\) & Predicted conditional severity \\
\(W\) & Set of evaluation windows \\
\(K\) & Number of counties staged per window \\
\(K_{t}\) & Number of counties staged in window \(t\) \\
\(\Phi_{t}(K)\) & Captured outage burden in window \(t\) \\
\(\alpha\) & Nominal miscoverage rate (0.10) \\
\(E_{i}\) & Nonconformity score for calibration point \(i\) \\
\(\widehat{q}\) & Conformal offset \\
\(\kappa\) & Mondrian taxonomy function \\
\(r\) & Cost--loss ratio, \(C_{prot}/L_{loss}\) \\
\(\theta\) & Probability threshold for a binary staging decision \\
\(h,\ f\) & Joint frequencies of hits and false alarms \\
\(\overline{s}\) & Climatological event frequency \\
\(V\) & Potential economic value \\
\(\rho\) & Fraction of requested crews secured \\
\(A_{C}\) & Avoided customer-interruption cost \\
\(H_{s}\) & Customer-hours of observed outage in staged counties \\
\(\beta\) & Mean demand per customer (1.2 kW) \\
\(\eta\) & Fraction of outage duration removed by pre-staging \\
\(\mathit{VoLL}\) & Value of lost load (USD/MWh) \\
\(D\) & Total staging commitment (county-days) \\
\(d\) & Days crews are held per commitment \\
\(m\) & Crews assigned per staged county \\
\(\nu_{j}\) & Across-year variation ratio of land-surface feature \(j\) \\
\end{longtable}
}

\subsection{Study domain and period}

We modelled electricity outages caused by storms at county resolution across the contiguous United States from 1 January 2022 to 31 December 2024. The county is the reporting unit of the outage record (Section 2.2), and 6 h windows match the time step at which the AI weather forecasts (Section 2.4) are archived. After coverage screening (Section 2.2), the analysis grid contains 2,491 counties and 4,384 time windows, giving 10.9 million county--window observations before unreported intervals were removed. The study period is set by the availability of the forecast archive, since the outage record extends further back. Earlier outage years are also unsuitable because of poor reporting coverage (Section 2.2). Table 1 summarises the data sources. Fig. 1 summarises the workflow, from the data sources through target and feature construction and the two prediction models to the evaluation of crew staging decisions.

\begin{figure}[!htbp]
\centering
\includegraphics[width=\linewidth,height=0.62\textheight,keepaspectratio]{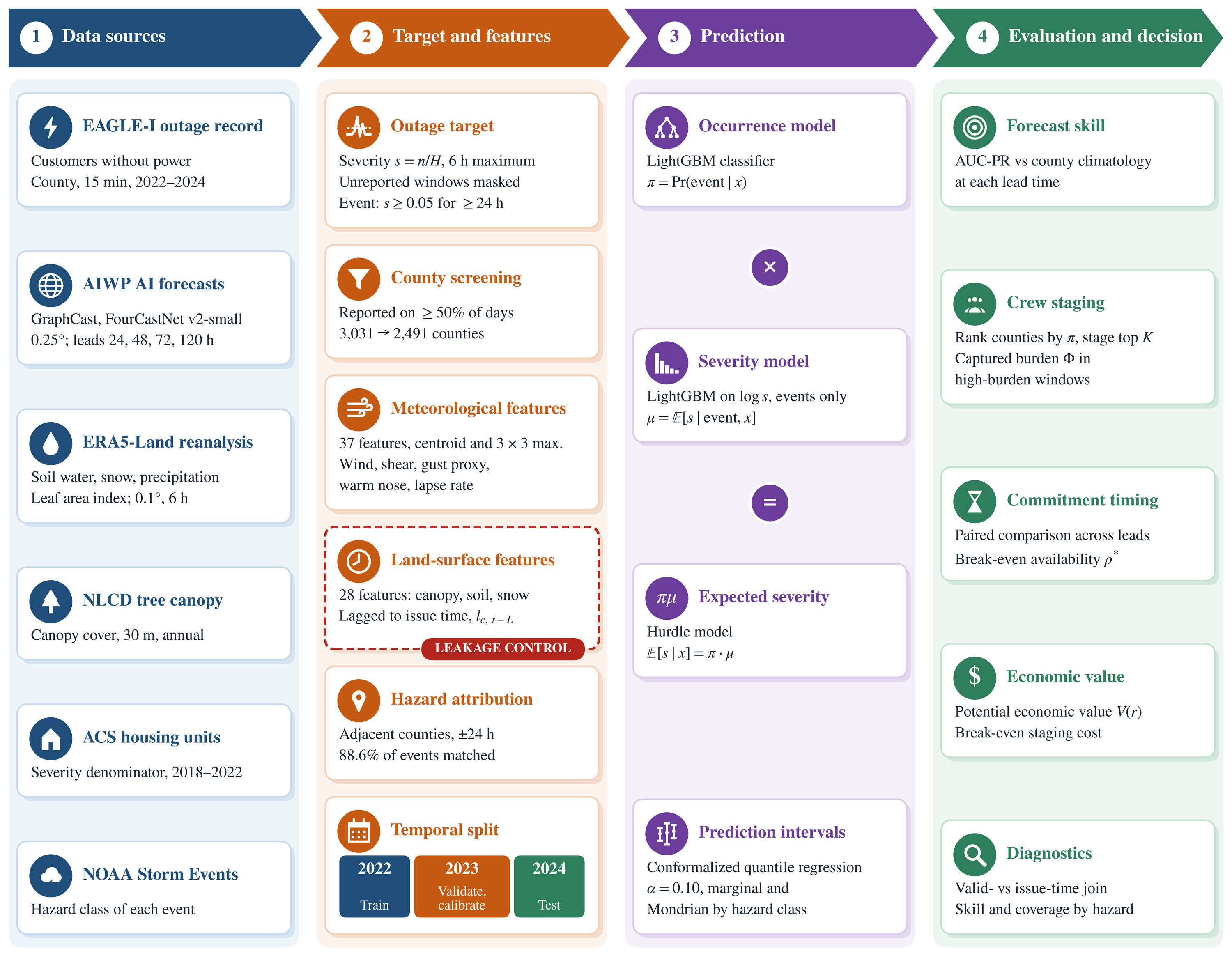}
\caption{\textbf{Study framework.} Outage records, AI weather forecasts, land surface reanalysis, canopy, housing and storm-report data (1) are combined into a county-level target and predictors, with land surface features lagged to the forecast issue time (2). Expected severity is the product of an occurrence probability and a conditional severity, and conformalized quantile regression gives severity intervals (3). Counties are ranked by predicted occurrence to evaluate crew staging, commitment timing and economic value (4).}
\label{fig:1}
\end{figure}

\begin{table}[!htbp]
\centering
\small
\caption{\textbf{Data sources.} The outage record sets the county and 15-minute native resolution of the target, and the forecast archive sets the 6-hourly analysis window. The severity denominator is drawn from a period preceding the study window so that it cannot encode information about the outcomes being predicted.}
\label{tab:1}
\begin{tabular}{@{}>{\raggedright\arraybackslash}p{\dimexpr 0.2621\linewidth-2\tabcolsep\relax}>{\raggedright\arraybackslash}p{\dimexpr 0.2236\linewidth-2\tabcolsep\relax}>{\raggedright\arraybackslash}p{\dimexpr 0.2036\linewidth-2\tabcolsep\relax}>{\raggedright\arraybackslash}p{\dimexpr 0.1553\linewidth-2\tabcolsep\relax}>{\raggedright\arraybackslash}p{\dimexpr 0.1553\linewidth-2\tabcolsep\relax}@{}}
\toprule
Source & Variable & Resolution & Period & Role \\
\midrule
EAGLE-I & Customers without power & County, 15 min & 2022--2024 & Target \\
NOAA/CIRA AIWP & GraphCast, FourCastNet v2-small & 0.25\textdegree{}, 6 h & 2022--2024 & Meteorological predictors \\
ERA5-Land & Soil moisture, leaf area index, precipitation, snow & 0.1\textdegree{}, 6 h & 2022--2024 & Land-surface predictors \\
NLCD & Tree canopy cover & 30 m, annual & 2022--2024 & Land-surface predictors \\
NOAA Storm Events & Hazard class & County or forecast zone, event & 2022--2024 & Hazard attribution \\
ACS 5-year & Housing units & County & 2018--2022 & Severity denominator \\
\bottomrule
\end{tabular}
\end{table}

\subsection{Outage data and target construction}

Outage data were taken from the Environment for Analysis of Geo-Located Energy Information (EAGLE-I) system of Oak Ridge National Laboratory, which collects outage reports from participating utilities in near real time and publishes the number of customers without power in each county every 15 minutes. The 2023 and 2024 files use different field names, which were merged into a single field before the years were combined.

Outage counts were normalised by a county denominator so that severity is comparable between counties of different size. Three denominators were considered. The customer counts provided with EAGLE-I are inconsistent, since in 79 counties the reported outage count exceeds the reported number of customers at least once, by up to a factor of 12.5. The maximum outage observed in each county during the study period (Arora et al., 2026) avoids this problem, but for many counties this maximum occurs during a 2024 event. A denominator built this way would include information from the test year in the target and make the evaluation circular.

We therefore normalised by the number of housing units from the 2018--2022 American Community Survey 5-year estimates (U.S. Census Bureau, 2023). This quantity is independent of the outage record, constant over the study period and taken from earlier years, so it contains no information about the outcomes being predicted. Severity is defined as

\begin{equation}
s_{c,t} = \frac{n_{c,t}}{H_{c}}
\label{eq:1}
\end{equation}
\noindent where \(n_{c,t}\) is the number of customers without power in county \(c\) at time \(t\) and \(H_{c}\) is the number of housing units. Because housing units and utility customers are not identical, \(s_{c,t}\) exceeds one in 293 counties at some point in the record, mainly where commercial accounts and accounts with several meters are common. These values were kept rather than clipped, because clipping would distort the upper tail that the severity model is meant to represent.

The 15 min records were aggregated to 6 h windows by taking the maximum, which keeps the peak of short events that a mean would reduce. County--windows with no records were masked and excluded from the model, rather than set to zero or filled with the last observation, because these two alternatives create outages or absences of outage that the record does not support (Supplementary Text S1 and Table S1). As a result, 46.3\% of county--window cells are unobserved and are excluded from both training and evaluation.

Reporting completeness varies between counties and improves over time. Counties that reported on fewer than 50\% of days in the study period were excluded, which reduced the domain from 3,031 to 2,491 counties. Applied to earlier years, the same screen removes almost all counties. In 2014 the median county reported on 9.3\% of days and no county passed the threshold. The set of reporting counties first reached 97\% overlap with 2024 in 2020 and stayed at that level only from 2023. This is the main reason the analysis does not extend further back.

An outage event is defined as a continuous run of county--windows in which severity is at or above a threshold \(\tau\) for at least a duration \(\mathrm{\Delta}\), allowing short interruptions of up to \(g\). The event indicator is

\begin{equation}
y_{c,t} = 1\left[ s_{c,t} \geq \tau \right], \qquad \tau = 0.05
\label{eq:2}
\end{equation}

Runs shorter than \(\mathrm{\Delta} = 24\)h were discarded, and gaps below the threshold of up to \(g = 12\)h were merged into the surrounding run. The resulting event rate is 2.03\% of observed county--windows, rising from 1.93\% in 2022 to 2.27\% in 2024.

The threshold and duration were checked against four documented storms rather than tuned. Supplementary Table S2 reports the number of events, the number of counties affected and the duration distribution for thresholds from 0.005 to 0.100. The median duration stays at 30 h from 0.005 to 0.050 and rises to 36 h at 0.100. A threshold of 0.05 is therefore the lowest at which the detected events are not dominated by small background outages, and the choice is not finely tuned. Table 2 lists the duration and peak severity of seven counties during these storms.

\begin{table}[!htbp]
\centering
\small
\caption{\textbf{Event-definition validation against documented storms.} Duration and peak severity are those recovered by the event definition of Section 2.2 for seven counties affected by four storms in the test year. Severities above one occur where commercial accounts and accounts with several meters make housing units an undercount of connected customers (Section 2.2). The peak severity for Harris County during Hurricane Beryl is consistent with the loss of approximately 88\% of customers reported independently by the local utility.}
\label{tab:2}
\begin{tabular}{@{}>{\raggedright\arraybackslash}p{\dimexpr 0.2788\linewidth-2\tabcolsep\relax}>{\raggedright\arraybackslash}p{\dimexpr 0.2981\linewidth-2\tabcolsep\relax}>{\raggedright\arraybackslash}p{\dimexpr 0.2702\linewidth-2\tabcolsep\relax}>{\raggedright\arraybackslash}p{\dimexpr 0.1529\linewidth-2\tabcolsep\relax}@{}}
\toprule
Storm & County & Duration (h) & Peak severity \\
\midrule
Beryl & Harris, TX & 204 & 0.90 \\
Beryl & Fort Bend, TX & 150 & 0.84 \\
Helene & Greenville, SC & 252 & 1.18 \\
Helene & Buncombe, NC & 420 & 1.72 \\
Milton & Hillsborough, FL & 174 & 0.86 \\
Milton & Pinellas, FL & 150 & 0.87 \\
May derecho & Harris, TX & 144 & 0.27 \\
\bottomrule
\end{tabular}
\end{table}

\subsection{Hazard attribution}

Each event was assigned to a hazard class using the NOAA Storm Events Database (NOAA National Centers for Environmental Information, 2025). Hazard classes are needed for the stratified evaluation (Section 2.8) and for calibration by hazard class (Section 2.7). Two features of this database required care. First, the county code field contains a forecast zone identifier when an event is recorded by zone. Treating the two codes as the same returns no ice storm events nationally, so the National Weather Service table that links zones to counties was used. Second, storm reports record damage where it was observed, not where the hazard occurred, so a convective system crossing several counties may produce reports in only some of them. Widening the time window did not solve this. Matching within the same county at \textpm{}48 h recovered few of the unmatched events, whereas allowing a match to an adjacent county at \textpm{}12 h recovered many more. The problem is therefore spatial rather than temporal. We adopted matching to adjacent counties at \textpm{}24 h, which gives a final match rate of 88.6\%. This assumes that a report in an adjacent county applies, which is reasonable for a squall line but less so for a local event. Section 3.5 reports the effect of restricting matches to the same county where it affects a result.

\subsection{Meteorological features}

Meteorological predictors were taken from the NOAA/CIRA Artificial Intelligence Weather Prediction (AIWP) archive, which provides reforecasts from operational AI global models on a 0.25\textdegree{} grid (Radford et al., 2025). Two models were selected based on the variables they provide. FourCastNet v2-small (Bonev et al., 2023) is the only archived model with 100 m wind, and GraphCast (Lam et al., 2023) is the only one with precipitation. None of the models provides wind gust, so a gust proxy was derived as described below. A third model, Aurora, is available only from 2025 onward and is outside the outage record.

Forecasts were used at lead times \(L \in \left\{ 24,\ 48,\ 72,\ 120 \right\}\ \)h. GraphCast output begins on 31 December 2021, so windows in early 2022 use FourCastNet only. A binary indicator marks the windows for which GraphCast is available, so that the model can distinguish the two data regimes and does not treat missing fields as physical values.

For each county centroid and forecast initialisation, we extracted surface fields (wind components at 10 and 100 m, 2 m temperature, surface pressure and mean sea level pressure) and fields at 850 and 700 hPa (temperature, relative humidity and wind). From these we derived wind speed at both heights, the wind shear between 10 and 100 m, the lapse rate between 850 and 700 hPa, a warm nose indicator and a gust proxy. The warm nose indicator flags an 850 hPa temperature above freezing with a surface temperature below freezing, which is the typical signature of freezing rain, and partly replaces the precipitation type field that the archive does not provide. The gust proxy adds a fixed fraction of the shear to the 10 m wind, \(g = w_{10} + c(w_{100} - w_{10})\), with \(c = 0.5\).

Because damaging wind is local, a county centroid may miss the strongest winds. Wind and gust quantities were therefore extracted both at the centroid and as the maximum over the 3\texttimes{}3 grid cells around it, assuming that county damage follows the most severe cell rather than the average. The meteorological vector \(m_{c,t}^{(L)}\) contains 37 features.

\subsection{Land surface features and the timing of information}

Land surface predictors come from two sources. Tree canopy cover was taken from the National Land Cover Database (NLCD) canopy product at 30 m resolution (Dewitz, 2024), using the product for each of 2022, 2023 and 2024. County statistics were computed over county polygons in an equal area projection, so that the unweighted mean of the pixels is an unbiased estimate of mean cover. Six canopy descriptors were kept, namely the mean, standard deviation and 90th percentile of cover, the fractions of pixels above 60\% and above 20\% cover, and the fraction between 20\% and 60\%. The last one represents forest edges, where wind loading on individual trees is highest and falling trees can reach power lines.

Soil, vegetation and snow conditions were taken from the ERA5-Land reanalysis (Hersbach et al., 2020; Muñoz-Sabater et al., 2021) at 0.1\textdegree{} resolution every 6 h. These include volumetric soil water at two depths, leaf area index for high and low vegetation, total precipitation and, for winter months, snow depth, snow water equivalent, snow density, snowfall, snow layer temperature and 2 m dewpoint. Derived quantities include antecedent precipitation over 24 and 72 h, a soil moisture anomaly relative to each county's monthly climatology, a dimensionless saturation index, and a snow load equal to snow water equivalent multiplied by snow density, since dry snow imposes a much smaller load than the same depth of wet snow. The land surface vector \(l_{c,t}\) contains 28 features.

Land surface predictors are often described as observations, but not all of them change from year to year. Supplementary Text S2 measures this for each feature. The canopy descriptors change very little between years, as expected for an annual product, and so does leaf area index, because ERA5 prescribes it from a monthly climatology instead of assimilating satellite retrievals. Soil moisture, precipitation and snow change with actual conditions, and we refer to them as prognostic. We therefore treat canopy and leaf area index as climatological features and report them separately from the prognostic features throughout.

A forecast issued at time \(t - L\ \)and valid at time \(t\) can only use land surface conditions known at or before the issue time. Evaluating land surface predictors at the valid time, which is what happens when reanalysis data are joined to the target on a shared time index, uses information that no operational forecaster has. We therefore define the land surface vector used throughout as

\begin{equation}
\widehat{l}_{c,t}^{(L)} = l_{c,t-L}
\label{eq:3}
\end{equation}
\noindent so that a prediction valid at \(t\) with lead \(L\) uses the land surface state at the time the forecast was issued. The difference is small for slowly varying quantities but large for quantities computed over rolling windows. For example, antecedent precipitation accumulated over the 72 h before the valid time includes precipitation that falls during the forecast period. Section 2.9 describes how this effect was tested, and Section 3.2 reports the results.

The full predictor vector is \(x_{c,t}^{(L)} = \left( m_{c,t}^{(L)}\ \ ,\ {\widehat{l}}_{c,t}^{(L)}\ \ ,\ z_{t} \right)\ \)where \(z_{t}\) is the data regime indicator, and contains 66 features. Two candidate predictors were removed after inspection. The number of canopy pixels in each county measures county area rather than vegetation, and the fraction of masked windows in each county measures reporting behaviour rather than weather. Both ranked high in preliminary feature importance, which is why they were removed.

\subsection{Predictive model}

The target contains many zeros and has a long upper tail. In total, 98.0\% of observed county--windows contain no event, and severities during events range over two orders of magnitude. A single regression model for severity minimises its error by predicting values close to zero everywhere, which gives a low error but no useful information. Hurdle models have been used for hurricane outage counts for this reason (Guikema and Quiring, 2012; Shashaani et al., 2018). We therefore split expected severity into an occurrence probability and a conditional magnitude,

\begin{equation}
\mathbb{E}\left[ s_{c,t} \mid x_{c,t}^{(L)} \right] = \underbrace{\Pr\left( y_{c,t} = 1 \mid x_{c,t}^{(L)} \right)}_{\pi_{c,t}} \cdot \underbrace{\mathbb{E}\left[ s_{c,t} \mid y_{c,t} = 1,\ x_{c,t}^{(L)} \right]}_{\mu_{c,t}}
\label{eq:4}
\end{equation}
\noindent and estimated the two parts separately. The occurrence part is a binary classifier trained on all observed county--windows, and the severity part is a regression for \(\log s_{c,t}\) trained on event windows only. Severity skill is therefore always conditional on occurrence, and a conditional coefficient of determination cannot be compared with one computed over all windows.

Both parts use gradient boosting decision trees implemented in LightGBM (Ke et al., 2017), with up to 1,000 estimators, a learning rate of 0.05, 63 leaves, row and feature sampling fractions of 0.8, an L2 penalty of 1.0, and a minimum of 100 observations per leaf for occurrence and 50 for severity. Training stopped when the average precision on the validation year did not improve for 80 rounds. The hazard-stratified skill (Table 8), the severity metrics and the wind comparison (Table 3) were computed with up to 2,000 estimators and a stopping window of 100 rounds. Missing values were handled directly by the model rather than imputed, which is important because 46.3\% of county--windows are unobserved.

Tree models were chosen after a comparison with a regularised linear model and a multilayer perceptron, with and without lagged features and features from adjacent counties (Supplementary Text S3 and Table S3). The tree model outperformed the neural network on the same features, consistent with benchmark evidence that tree ensembles often outperform neural networks on tabular data (Grinsztajn et al., 2022). For this dataset and setup the tree model is therefore the better choice, although the comparison does not establish whether another architecture could extract more from the same predictors. Adding lagged features raised its skill by 5.4\%. We kept the tree model without these structured features because the decision results in Section 3.1 depend on the top of the county ranking rather than on overall skill, and because tree models handle missing windows without imputation.

\subsection{Predictive uncertainty}

A staging decision needs a reliable interval, not only a point prediction of severity. Interval predictions are still rare in outage prediction, with quantile regression forests (He et al., 2017) and probabilistic hurricane outage models (Arora and Ceferino, 2023) as the main exceptions. We constructed prediction intervals by conformalized quantile regression (Romano et al., 2019), which requires no assumption about the error distribution. Two quantile regressions were fitted on the training year at levels \(\alpha/2\) and \(1 - \alpha/2\) , giving \({\widehat{q}}_{lo}(x_{i})\) and \({\widehat{q}}_{hi}(x_{i})\) for \(\alpha = 0.10\). On a separate calibration set of \(n\) event windows, we computed the nonconformity score

\begin{equation}
E_{i} = \max\left\{ \widehat{q}_{lo}(x_{i}) - \log s_{i},\ \log s_{i} - \widehat{q}_{hi}(x_{i}) \right\}
\label{eq:5}
\end{equation}
\noindent and took \(\widehat{q}\) as the \(\left\lceil (n + 1)(1 - \alpha) \right\rceil/n\) empirical quantile of \(\left\{ E_{i} \right\}\). The prediction interval for a new point is

\begin{equation}
C(x) = \left[ \widehat{q}_{lo}(x) - \widehat{q},\ \widehat{q}_{hi}(x) + \widehat{q} \right]
\label{eq:6}
\end{equation}
\noindent which has marginal coverage of at least \(1 - \alpha\) if calibration and test points are exchangeable.

Marginal coverage is a weak guarantee when the test data differ from the calibration data in a known way. We therefore also computed a Mondrian version, in which a function \(\kappa\) assigns each point to a hazard class and a separate offset is calibrated within each class (Vovk, 2013),

\begin{equation}
\widehat{q}^{(k)} = \mathrm{Quantile}\left( \left\{ E_{i} : \kappa(i) = k \right\},\ \frac{\left\lceil (n_{k} + 1)(1 - \alpha) \right\rceil}{n_{k}} \right)
\label{eq:7}
\end{equation}

This gives coverage of at least \(1 - \alpha\) within every class with enough calibration points. Coverage is reported only for classes with enough test events for a meaningful estimate, so ice storms, with 18 test events, were excluded. Hazard class is assigned from storm reports after the event. The Mondrian result is therefore a diagnostic that shows the coverage failure is conditional rather than marginal. An operational version would need a hazard class derived from the forecast itself.

The calibration set must be separate from the data used to fit the model. In our design, the 2023 data serve three roles, namely early stopping, model selection and conformal calibration. The quantile regressions therefore use early stopping on the same year that provides the nonconformity scores. These scores are not fully independent of the fitted model, and the marginal coverage of 0.861 is likely optimistic. A fourth data partition would remove this problem, but the calibration sample would then be too small for the analysis by hazard class in Section 3.4. Because this bias overstates coverage, the conditional failure reported in Section 3.4 is, if anything, understated. To show why a separate calibration set matters, we also report a version in which the training year is added to the calibration set. Residuals on training data are smaller than on new data, so this reduces \(\widehat{q}\) and lowers coverage (Section 3.4).

\subsection{Evaluation protocol}

The three years were assigned to separate roles, with 2022 used for training, 2023 for early stopping, model selection and conformal calibration, and 2024 for testing. The test year was not used for training or selection. A temporal split was used instead of a random split because county--windows are strongly dependent in space and time, and a random split would place windows from the same storm in both training and test data. The 2024 data contain about half as many forecast initialisations as 2023, two per day instead of four, so the test sample is smaller than a full year and confidence intervals are wider.

Forecast initialisations are missing near the start and end of the archive, so each lead time covers a slightly different set of valid times. Comparing lead times on their own samples would therefore compare different storms rather than different forecast quality. All comparisons across lead times were restricted to the 4,721,771 county--windows available at every lead.

Occurrence performance is measured by the area under the precision--recall curve (AUC-PR). At a base rate near 2\%, this metric is preferred to the area under the receiver operating characteristic curve, which is dominated by true negatives. Skill is compared with a climatological reference, the historical event rate of each county in the training year. Evaluation by hazard class needs care, because hazard labels come from the events. Evaluating within a hazard class conditions on the outcome, since almost all windows in the class are events, and the result cannot be read as skill. We instead compared event windows of each hazard with control windows from the same counties and calendar months. We report lift over the local base rate together with a normalised skill score, because lift is bounded by the reciprocal of the base rate and cannot be compared directly between classes with very different base rates.

Windows within the same storm share the same forecast error and are not independent. Confidence intervals were therefore computed by a block bootstrap that resamples storm weeks rather than individual windows. The high burden windows of the test year fall in 18 storm weeks, with a median of three windows per week and a maximum of 14. These blocks, not the windows, set the effective sample size for every interval on captured burden. This is why these intervals are wide compared with the skill intervals in Section 3.1, which are estimated over 4.7 million county--windows. A block bootstrap with so few blocks tends to produce intervals that are too narrow, so these intervals should be read as approximate. The significant differences are better supported by their consistency across the onset and hazard analyses in Section 3.1 than by the coverage of any single interval. Resampling windows independently would make the intervals about three times narrower in these data.

Lead times were compared using paired differences computed within each bootstrap sample, not by comparing separate intervals. Because every lead is evaluated on the same resampled storm weeks, the large variation caused by which storms enter each sample affects both leads equally and cancels in the difference. Overlapping intervals are not a test of a difference, and the paired comparison is much more sensitive.

Feature groups were defined in a single manifest file with a content hash, and every result reported here was produced with that manifest. The group definitions, the two excluded predictors (Section 2.5) and the data partitions are therefore the same in all analyses. The manifest and the scripts that use it are provided with the data described in the Data Availability statement.

\subsection{Isolating the effect of information timing}

The effect of evaluating land surface predictors at the valid time instead of the issue time was measured with three tests. First, for each land surface feature, we computed the Pearson correlation between its values at valid time and at issue time as a function of lead time. This shows directly how much of the value at valid time is already known at issue time. Second, we refitted the occurrence model under both conventions at every lead time and compared the skill, keeping everything else the same. Third, we used a permutation test that keeps county identity but removes timing information. The land surface vector of each county--window was replaced by that of the same county on a randomly chosen different date. If land surface skill comes from conditions before the storm, it should disappear under this permutation. If it comes from fixed county characteristics, it should remain.

\subsection{Decision analysis}

Before a storm, a utility must decide how many crews to request through mutual assistance and where to place them. Mutual assistance rates are agreed in advance and do not depend on when a request is made. The cost of a late request is therefore not a higher price but fewer crews, because utilities compete for the same limited pool. The decision also fixes crew locations, because staging sites need lodging, fuel and equipment arranged in advance, and crews travelling from distant utilities cannot be redirected once they have left.

Within an evaluation window \(t\), counties were ranked by predicted occurrence probability \(\pi_{c,t}\) and crews were staged in the \(K\) counties with the highest probability, with ties broken arbitrarily. The captured outage burden is the share of the observed severity in that window that falls within the staged set,

\begin{equation}
\Phi_{t}(K) = \frac{\sum_{c \in S_{t}(K)} s_{c,t}}{\sum_{c} s_{c,t}}, \qquad S_{t}(K) = \left\{ c : \mathrm{rank}_{c}\, \pi_{c,t} \leq K \right\}
\label{eq:8}
\end{equation}

Counties were ranked by predicted occurrence rather than predicted severity. The severity model therefore does not affect the staging decision, and its limitations do not affect the results in Section 3.1. \(\Phi\) is a verification metric rather than a dispatch model. It measures whether a forecast concentrates observed burden into a small set of counties, which any useful staging rule requires, and not how a particular utility would allocate crews within its own territory and contracts.

Staging decisions are made before significant events, not for routine outages. Averaging over all windows with any outage makes the comparison uninformative. Even a model with perfect foresight captures only 36\% of burden when staging ten counties across all windows, because widespread small outages cannot be concentrated by any ranking. We therefore report \(\Phi\) for the highest decile of windows by observed burden, together with the result for all windows. In the test year this decile contains 72 windows. One of them is not available at every lead time, so comparisons between leads, which are paired by window, use the 71 windows common to all four leads.

Because events are defined as continuous runs above the severity threshold, a window may fall in the restoration phase after a storm has passed, when staging can no longer help. Within the high burden windows, the median event window lies 12 h after onset, with an interquartile range of 0 to 36 h, and 57.3\% of observed burden falls within 24 h of onset. Most of these windows therefore occur during the storm. Section 3.1 also reports the decision results with burden restricted to counties near onset.

Because the high burden windows are selected using observed burden, we also tested whether they can be identified in advance. Summing the model's county predictions within each window and ranking windows by this total identifies the highest decile windows with an area under the receiver operating characteristic curve of 0.785 nationally, compared with 0.688 for a climatological reference. Both steps, first identifying an important window and then allocating crews within it, can therefore be done before the event. Section 3.1 reports captured burden computed on the windows flagged by the model instead of the observed highest decile.

Let \(\rho \in (0,\ 1\rbrack\) be the fraction of requested crews that a utility actually receives, which decreases as the request is delayed. Reduced availability was modelled by staging the \(\rho K\) counties ranked highest instead of \(K\). This assumes that crews are removed from the lowest ranked staged counties first, which is the most favourable case, and a utility that loses crews from random locations would capture less. The form of \(\rho(L)\) cannot be observed in any public record. Instead of assuming it, we report the break-even ratio, the value of \(\rho\) at which committing at 24 h with fewer crews equals committing at 72 h with all crews,

\begin{equation}
\rho^{*} : \quad \Phi\left( \rho^{*}K; L = 24 \right) = \Phi(K; L = 72)
\label{eq:9}
\end{equation}

A utility can compare this single value with its own mutual assistance experience, and it requires no assumption about how availability declines.

To express forecast quality without assuming a specific cost, we computed potential economic value within the cost--loss framework (Murphy, 1977; Richardson, 2000; Wilks, 2001). A user who pays a protection cost \(C_{prot}\) to act, and suffers a loss \(L_{loss}\) if an event occurs without protection, is characterised by the ratio \(r = \frac{C_{prot}}{L_{loss}}\). For a binary decision taken at threshold \(\theta\), with \(h\) and \(f\) the joint frequencies of hits and false alarms over all forecast-outcome pairs, and \(\overline{s}\) the climatological event frequency, value relative to climatology is

\begin{equation}
V(r) = \max \frac{\min\left( r, \overline{s} \right) - \left[ r(h + f) + \left( \overline{s} - h \right) \right]}{\min\left( r, \overline{s} \right) - r\overline{s}}
\label{eq:10}
\end{equation}
\noindent which equals one for a perfect forecast and zero for a forecast no better than climatology. Reporting \(V\) across \(r\) allows each utility to find its position on the horizontal axis from its own costs, without assuming a value of lost load.

To express captured burden in monetary terms, it is converted to avoided customer-interruption cost as

\begin{equation}
A_{C} = H_{s} \cdot \beta \cdot \eta \cdot \mathit{VoLL}
\label{eq:11}
\end{equation}
\noindent where \(H_{s}\) is the customer-hours of observed outage falling in staged counties, \(\beta = 1.2\) kW is the mean demand per customer from national residential consumption statistics, varied from 0.8 to 2.0 kW in the sensitivity analysis of Fig. 6d, \(\eta\) is the fraction of outage duration removed by staging crews in advance, and \(VoLL\) is the value of lost load. Two of these quantities cannot be measured from our data and were varied instead of fixed. \(VoLL\) was varied from 500 to 5,000 USD/MWh, which covers published estimates and the current administrative value used by one large system operator. The restoration acceleration factor \(\eta\) was varied from 0.10 to 0.50. Staging crews in advance does not prevent outages but shortens restoration, so we report avoided customer hours under a stated acceleration rather than avoided outages. Crew cost was also not assumed, because published daily rates for storm crews are not available in citable form. We instead report the break-even staging cost per county-day for each combination of \(VoLL\) and \(\eta\).

\section{Results and discussion}

\subsection{Timing of crew commitment}

Occurrence skill, measured by AUC-PR on the 2024 test year, decreases steadily with lead time, from 0.1207 at 24 h to 0.0646 at 120 h for the full set of predictors, and from 0.1027 to 0.0625 for meteorology alone. The AI forecasts underestimate wind speed. Compared with ERA5 at the same 0.25\textdegree{} resolution, AIWP 10 m wind has a median ratio of 0.915 across 6.8 million county--windows, falling to 0.901 in the highest decile of observed wind. The forecasts therefore underestimate wind throughout the distribution, and slightly more at high wind speeds. At every lead time the model remains well above the county climatological reference of 0.047, so even a forecast made five days ahead provides information that a utility cannot obtain from history alone.

Absolute skill is modest. An AUC-PR of 0.12 against a base rate of 2.27\% is 5.3 times better than random selection, but most of the variation remains unexplained. Part of this reflects the predictors available. The wind comparison below shows that the AI forecasts lack a gust field, and adding one raises the skill of wind-only models by about a third (Table 3). Whether a different model class could extract more from the same predictors is not established here (Section 2.6).

The effect of wind forecast error can be separated from the effect of missing variables. Replacing AIWP forecast wind with ERA5 analysis wind at valid time, using the same three quantities, raises occurrence skill from 0.0463 to 0.0530, a difference of +0.0071 with a 95\% interval of {[}+0.0013, +0.0140{]} (Table 3). Adding the ERA5 instantaneous gust field, which the AIWP archive does not provide at any lead time, raises skill much more, by +0.0176 {[}+0.0110, +0.0237{]} with analysis wind and by +0.0169 {[}+0.0113, +0.0219{]} with forecast wind. These two gains are similar, which indicates that the gust field adds information regardless of forecast quality.

For wind, the main limitation is therefore the missing gust field rather than errors in the fields that are available. Perfect wind forecasts would improve skill by about 15\%, whereas a gust observation would improve it by about 33\%, and better forecast models cannot provide it. The three analysis fields also outperform the 17 wind features of the full model, so these engineered features already extract most of what the archived wind can provide. This comparison has three limits. First, ERA5 analysis wind is an upper bound rather than an achievable target, because the reanalysis includes observations taken during the event. Second, it covers wind only. Thermodynamic fields have a larger share of model attribution (Section 3.3), and their forecast error is not bounded here. Both wind specifications are also close to the county climatological reference of 0.047, so the percentage gains are computed on a small base and do not apply to the full model. Third, an archived gust field would itself have forecast errors that grow with lead time, which a substitution at valid time cannot show.

\begin{table}[!htbp]
\centering
\small
\caption{\textbf{Upper bound on wind skill from analysis data.} The matched specifications use 10 m wind, 100 m wind and the shear between them at the county centroid, from AIWP forecasts or ERA5 analysis. The gust field is the ERA5 instantaneous 10 m gust, which the AIWP archive does not provide. ERA5 analysis at valid time is an upper bound on any forecast of these fields, not an achievable skill.}
\label{tab:3}
\begin{tabular}{@{}>{\raggedright\arraybackslash}p{\dimexpr 0.3689\linewidth-2\tabcolsep\relax}>{\raggedright\arraybackslash}p{\dimexpr 0.4272\linewidth-2\tabcolsep\relax}>{\raggedright\arraybackslash}p{\dimexpr 0.2039\linewidth-2\tabcolsep\relax}@{}}
\toprule
Wind source & Features & AUC-PR \\
\midrule
AIWP forecast, full & 17 & 0.0513 \\
AIWP forecast, matched & 3 & 0.0463 \\
ERA5 analysis, matched & 3 & 0.0530 \\
ERA5 analysis + gust & 4 & 0.0712 \\
AIWP forecast + gust & 18 & 0.0685 \\
\bottomrule
\end{tabular}
\end{table}

For a utility, the relevant quantity is not skill but the share of observed outage burden in the counties where crews are staged. Table 4 reports this captured burden by lead time and by the fraction of requested crews secured, and Fig. 2c shows it for the full range of staging fractions together with a perfect foresight benchmark. Staging crews in the 10\% of counties ranked highest during high burden windows captures 0.356 of observed burden at 24 h and 0.238 at 120 h. The perfect foresight benchmark captures 0.893 under the same rule, so a large gap remains at every lead time.

\begin{table}[!htbp]
\centering
\small
\caption{\textbf{Captured outage burden by commitment lead and crew availability.} Crews are staged in the 10\% of counties ranked highest during high burden windows of the 2024 test year. Values in brackets are 95\% intervals from a block bootstrap over storm weeks. The climatology and oracle columns are computed with all requested crews.}
\label{tab:4}
\begin{tabular}{@{}>{\raggedright\arraybackslash}p{\dimexpr 0.1312\linewidth-2\tabcolsep\relax}>{\raggedright\arraybackslash}p{\dimexpr 0.2514\linewidth-2\tabcolsep\relax}>{\raggedright\arraybackslash}p{\dimexpr 0.1270\linewidth-2\tabcolsep\relax}>{\raggedright\arraybackslash}p{\dimexpr 0.1346\linewidth-2\tabcolsep\relax}>{\raggedright\arraybackslash}p{\dimexpr 0.1260\linewidth-2\tabcolsep\relax}>{\raggedright\arraybackslash}p{\dimexpr 0.1347\linewidth-2\tabcolsep\relax}>{\raggedright\arraybackslash}p{\dimexpr 0.0951\linewidth-2\tabcolsep\relax}@{}}
\toprule
Lead (h) & 100\% crews & 80\% & 60\% & 40\% & Climatology & Oracle \\
\midrule
24 & 0.356 {[}0.264, 0.454{]} & 0.308 & 0.258 & 0.204 & 0.192 & 0.893 \\
48 & 0.327 {[}0.254, 0.414{]} & 0.285 & 0.240 & 0.187 & 0.192 & 0.893 \\
72 & 0.316 {[}0.260, 0.384{]} & 0.279 & 0.241 & 0.176 & 0.190 & 0.894 \\
120 & 0.238 {[}0.167, 0.312{]} & 0.198 & 0.158 & 0.116 & 0.194 & 0.893 \\
\bottomrule
\end{tabular}
\end{table}

The shape of this decline is the main result. Captured burden falls by 0.040 over the first three days and by a further 0.079 between three and five days, so most of the loss occurs at long lead times. Paired bootstrap differences confirm this (Table 5 and Fig. 2e). The difference between 72 and 120 h excludes zero, whereas the difference between 24 and 72 h does not. The difference between 24 and 48 h is significant, although the larger difference between 24 and 72 h is not. This is expected, because adjacent lead times share more forecast initialisations and therefore have narrower intervals.

\begin{table}[!htbp]
\centering
\small
\caption{\textbf{Paired differences in captured burden between commitment leads.} Computed within each bootstrap sample so that variation common to both leads cancels. Positive values mean the earlier lead captures more. A difference is significant when its 95\% interval excludes zero.}
\label{tab:5}
\begin{tabular}{@{}>{\raggedright\arraybackslash}p{\dimexpr 0.2913\linewidth-2\tabcolsep\relax}>{\raggedright\arraybackslash}p{\dimexpr 0.2524\linewidth-2\tabcolsep\relax}>{\raggedright\arraybackslash}p{\dimexpr 0.3010\linewidth-2\tabcolsep\relax}>{\raggedright\arraybackslash}p{\dimexpr 0.1553\linewidth-2\tabcolsep\relax}@{}}
\toprule
Comparison & Difference & 95\% interval & Excludes zero \\
\midrule
24 h vs 48 h & +0.029 & {[}+0.004, +0.065{]} & yes \\
48 h vs 72 h & +0.012 & {[}\textminus{}0.021, +0.054{]} & no \\
24 h vs 72 h & +0.040 & {[}\textminus{}0.006, +0.102{]} & no \\
72 h vs 120 h & +0.079 & {[}+0.036, +0.120{]} & yes \\
24 h vs 120 h & +0.120 & {[}+0.067, +0.187{]} & yes \\
\bottomrule
\end{tabular}
\end{table}

These results identify a boundary but not an optimum. Committing at five days costs significantly more than committing at three days, but the data do not show any significant ordering between 24 and 72 h. As a bound, committing at 72 h instead of 24 h costs at most 10.2 percentage points of captured burden at 95\% confidence, which is at most 29\% of the 0.356 captured at 24 h. On this pooled measure, a utility that prefers to commit at 48 or 72 h for logistical reasons would not lose a measurable share of captured burden. The onset analysis below shows that this conclusion is incomplete.

\begin{figure}[!htbp]
\centering
\includegraphics[width=\linewidth,height=0.62\textheight,keepaspectratio]{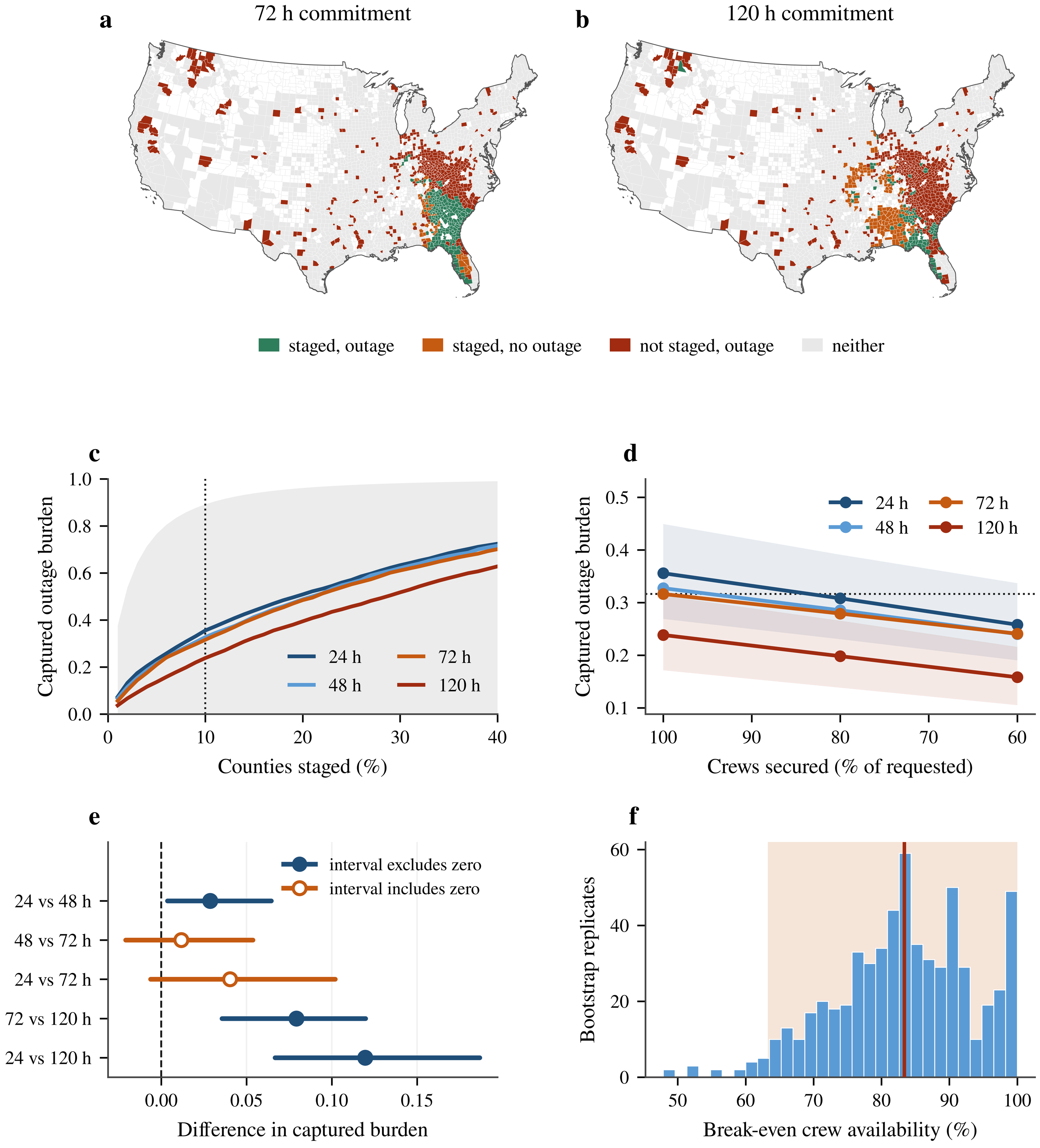}
\caption{\textbf{The crew commitment decision.} (a, b) Staging outcome for Hurricane Helene with commitment at 72 and 120 hours, with crews placed in the 10\% of counties ranked highest by predicted occurrence probability. Green counties were staged and had an outage, orange counties were staged without an outage, and red counties had an outage without being staged. Captured burden falls from 47.5\% at 72 hours to 18.7\% at 120 hours, and the loss is concentrated along the storm track through the southern Appalachians. Counties away from the track reflect unrelated outages elsewhere within the eight days of the evaluation window. Helene is shown as an illustration and is not the basis of any conclusion by hazard, since it contributes 10 of the 14 windows dominated by tropical cyclones in the test year. (c) Captured outage burden against the fraction of counties staged, by lead time, over high burden windows of the 2024 test year. The shaded envelope is a perfect foresight benchmark, and the dotted line marks the 10\% staging fraction used elsewhere. (d) Captured burden against the fraction of requested crews secured, with lead times coloured as in panel (c). The dotted horizontal line marks captured burden at 72 hours with all crews, which exceeds that at 24 hours with 60\% of crews. Shaded bands are 95\% intervals for the shortest and longest lead times. (e) Paired differences in captured burden between lead times, computed within each bootstrap sample so that variation common to both leads cancels. Blue lines with filled markers are intervals that exclude zero, and orange lines with open markers are intervals that include zero. (f) Bootstrap distribution of the break-even crew availability ratio at which committing at 24 hours equals committing at 72 hours with all crews. The vertical line is the point estimate of 83.4\% and the shaded region is its 95\% interval.}
\label{fig:2}
\end{figure}

The comparisons above assume the same number of crews at every lead time, which is not the decision a utility faces. Fig. 2d shows captured burden as a function of the fraction of requested crews secured. Committing at 72 h with all crews captures 0.316, more than committing at 24 h with 60\% of crews (0.258). The break-even ratio defined in Equation (9) is 83.4\%, with a 95\% interval of {[}63.3\%, 100\%{]} (Fig. 2f). Early commitment is therefore better unless a utility can expect to secure more than about 83\% of its requested crews one day before the storm. The interval is wide and reaches 100\%, so we cannot exclude that late commitment is always better for a utility that faces no competition for crews. The ratio is still more useful than an assumed availability curve, because it requires no assumed functional form and can be compared directly with a utility's own mutual assistance records.

The results above use windows selected by observed burden. Using instead the windows that the model itself would have flagged (Section 2.10) gives captured burden of 0.345 at 24 h, 0.325 at 48 h, 0.289 at 72 h and 0.243 at 120 h, compared with 0.356, 0.327, 0.316 and 0.238 for the observed highest decile windows. The two sets agree within 0.03 at every lead time, so the allocation results do not depend on selecting windows with hindsight.

Long events continue to contribute outage windows after the weather has cleared. To test whether the lead time result reflects forecasts made before the hazard rather than outages already in progress, we recomputed \(\Phi\) counting only counties within a fixed time after the onset of their event. Windows were ranked again by this restricted burden, and the highest decile was kept. Counties were still ranked by predicted occurrence, so the outcome does not enter the decision.

Table 6 reports the results. Captured burden is higher in all cases, because new outages are more concentrated in the counties ranked highest by the model than accumulated burden is. The columns have different denominators, so their levels cannot be compared. Within 24 h of onset, captured burden falls from 0.506 at 24 h to 0.347 at 120 h. The difference between 24 and 72 h is +0.075 with a 95\% interval of {[}+0.030, +0.121{]}, which excludes zero, whereas the corresponding unrestricted difference does not. The difference between 72 and 120 h remains significant for every onset window, at +0.084 {[}+0.044, +0.136{]} within 24 h of onset.

\begin{table}[!htbp]
\centering
\small
\caption{\textbf{Captured outage burden with burden restricted to counties near event onset.} Windows are ranked again by the restricted burden and the highest decile is kept, while counties are still ranked by predicted occurrence probability. Values in brackets are 95\% intervals from a block bootstrap over storm weeks. The columns have different denominators, so their levels cannot be compared.}
\label{tab:6}
\begin{tabular}{@{}>{\raggedright\arraybackslash}p{\dimexpr 0.2136\linewidth-2\tabcolsep\relax}>{\raggedright\arraybackslash}p{\dimexpr 0.3107\linewidth-2\tabcolsep\relax}>{\raggedright\arraybackslash}p{\dimexpr 0.2621\linewidth-2\tabcolsep\relax}>{\raggedright\arraybackslash}p{\dimexpr 0.2136\linewidth-2\tabcolsep\relax}@{}}
\toprule
Lead (h) & Within 0 h & Within 24 h & Within 48 h \\
\midrule
24 & 0.497 {[}0.442, 0.558{]} & 0.506 {[}0.437, 0.582{]} & 0.460 {[}0.410, 0.530{]} \\
48 & 0.453 {[}0.396, 0.519{]} & 0.465 {[}0.398, 0.538{]} & 0.426 {[}0.374, 0.492{]} \\
72 & 0.419 {[}0.359, 0.488{]} & 0.431 {[}0.366, 0.502{]} & 0.392 {[}0.340, 0.445{]} \\
120 & 0.331 {[}0.282, 0.389{]} & 0.347 {[}0.292, 0.409{]} & 0.309 {[}0.264, 0.359{]} \\
\bottomrule
\end{tabular}
\end{table}

The flat region between 24 and 72 h is therefore partly caused by burden in counties whose outages began before the commitment decision. For the burden that staging can still affect, captured burden declines steadily and the loss from waiting within the first three days is significant. We keep the unrestricted measure as the main result because it makes no assumption about which burden a crew can affect, and we report the restricted measure as a bound on how much of the flat region is caused by event duration.

Stratifying by hazard gives the same conclusion. Of the 71 high burden windows common to all four lead times, 57 are not tropical, and for these windows the difference between 24 and 72 h is significant at +0.062 {[}+0.015, +0.116{]}, whereas it is not significant in the pooled sample. The 14 tropical windows behave differently, but they are dominated by a single storm, and we draw no conclusion by hazard from them for the reasons given in Section 3.5. The flat region within the first three days is therefore not a general property of the forecast, and two independent divisions of the data, by event age and by hazard, agree on this.

Fig. 2a and 2b show the staging outcome for Hurricane Helene with commitment at 72 h and 120 h. Captured burden falls from 0.475 to 0.187, and the maps show why. At 72 h the staged counties follow the storm track through the southern Appalachians, where the largest burden occurred, whereas at 120 h the staged counties are spread out and most of the counties with the highest burden are missed. Helene is shown as an illustration, not as evidence. It contributes 10 of the 14 windows dominated by tropical cyclones in the test year, and Section 3.5 explains why no conclusion by hazard is drawn from it.

\subsection{Temporal leakage in antecedent land surface predictors}

Land surface predictors are usually joined to the forecast target on a shared time index, which evaluates them at the valid time. For a forecast issued \(L\) hours earlier, this uses information that no operational forecaster has (Fig. 3a). How serious this is depends on how much of the value at valid time is already known at issue time.

Table 7 and Fig. 3b report the correlation between the values at valid time and at issue time for each feature and lead time. Most of the soil moisture signal is preserved, with the correlation falling from 0.984 at 24 h to 0.926 at 120 h, and the same holds for snow load (0.998 to 0.976). Leaf area index is preserved almost perfectly, because it is prescribed from a climatology. Antecedent precipitation behaves very differently, with the correlation falling from 0.782 at 24 h to 0.053 at 120 h. The reason is that a 72 h accumulation window ending at the valid time hardly overlaps with the same window shifted back by 120 h. At long lead times this feature is therefore not a measure of antecedent wetness but a partial observation of the storm being forecast.

\begin{table}[!htbp]
\centering
\small
\caption{\textbf{Recoverability and time variation of land-surface predictors.} Correlation between the values of each land surface predictor at valid time and at issue time, by lead time, together with the variation ratio between years (Equation S1 in the Supplementary Material). A ratio near zero indicates a prescribed climatology rather than actual conditions. Leaf area index and canopy cover fall in this category, so neither represents the actual vegetation state of a given season. Canopy cover is an annual product, so its values at valid time and issue time are identical and no correlation is reported (n/a).}
\label{tab:7}
\begin{tabular}{@{}>{\raggedright\arraybackslash}p{\dimexpr 0.3069\linewidth-2\tabcolsep\relax}>{\raggedright\arraybackslash}p{\dimexpr 0.1188\linewidth-2\tabcolsep\relax}>{\raggedright\arraybackslash}p{\dimexpr 0.1188\linewidth-2\tabcolsep\relax}>{\raggedright\arraybackslash}p{\dimexpr 0.1188\linewidth-2\tabcolsep\relax}>{\raggedright\arraybackslash}p{\dimexpr 0.1297\linewidth-2\tabcolsep\relax}>{\raggedright\arraybackslash}p{\dimexpr 0.2069\linewidth-2\tabcolsep\relax}@{}}
\toprule
Feature & 24 h & 48 h & 72 h & 120 h & Variation ratio \(\mathbf{\nu}\) \\
\midrule
Antecedent precipitation (72 h) & 0.782 & 0.449 & 0.171 & 0.053 & 0.597 \\
Soil moisture (layer 2) & 0.984 & 0.968 & 0.953 & 0.926 & 0.325 \\
Snow load & 0.998 & 0.993 & 0.988 & 0.976 & 0.078 \\
Leaf area index (high vegetation) & 1.000 & 1.000 & 0.999 & 0.998 & 0.005 \\
Canopy cover & n/a & n/a & n/a & n/a & 0.007 \\
\bottomrule
\end{tabular}
\end{table}

Refitting the occurrence model under both conventions shows the size of the effect. When land surface predictors are joined at issue time instead of valid time, their skill falls from 0.0775 to 0.0452 at 24 h and from 0.0779 to 0.0369 at 120 h, a reduction of 42\% to 53\% depending on lead time (Fig. 3c). The correction also reverses a conclusion. With joining at valid time, land surface predictors appear to outperform forecast meteorology at 120 h (0.0779 against 0.0625), whereas with joining at issue time they do not (0.0369 against 0.0625). The corrected values are at or below the county climatological reference of 0.047 at every lead time. When evaluated with the information available to an operational forecaster, land surface conditions alone therefore do not predict outage occurrence at county scale.

The permutation test shows what remains after the correction. Replacing the land surface vector of each county--window with that of the same county on a randomly chosen different date keeps county identity but removes timing information. This removes 58.2\% of the land surface skill with joining at valid time, but only 22.6\% with joining at issue time (Fig. 3d). Most of the skill that remains after the timing correction therefore comes from fixed county characteristics rather than from conditions before the storm, which agrees with the attribution analysis in Section 3.3.

Antecedent conditions computed over rolling windows are widely used as predictors in impact forecasting, including flood, landslide and wildfire applications as well as outage prediction. Whenever the accumulation window ends at the valid time of the prediction rather than its issue time, the same contamination occurs, and it grows with lead time. This is a form of the temporal leakage described in broader reviews of machine learning practice (Kapoor and Narayanan, 2023). We do not claim to have found it in a published outage model. The leaky configuration evaluated here was built on purpose, by joining on a shared time index as a simple implementation would, so that its size could be measured. The point is that this join is the default result of the most natural data preparation step, not that any particular study has made this error.

The diagnostic is simple. Computing the correlation between the values of a feature at valid time and at issue time across the lead times of interest, as in Fig. 3b, identifies the affected features immediately and does not require fitting a model.

Correcting the convention does not make the predictor operational. Evaluation at issue time is an upper bound on the information that can legitimately be used, not an operational specification. ERA5-Land is published several days after real time, so a forecaster at issue time would have neither the value at valid time nor the value at issue time, but an earlier one. This does not affect the diagnostic, which compares the two conventions, nor the reversal reported above, since the values at issue time are an upper bound and the corrected skill is already below county climatology. An operational system would need a land surface product with a short delay or a proxy derived from the forecast.

\begin{figure}[!htbp]
\centering
\includegraphics[width=\linewidth,height=0.62\textheight,keepaspectratio]{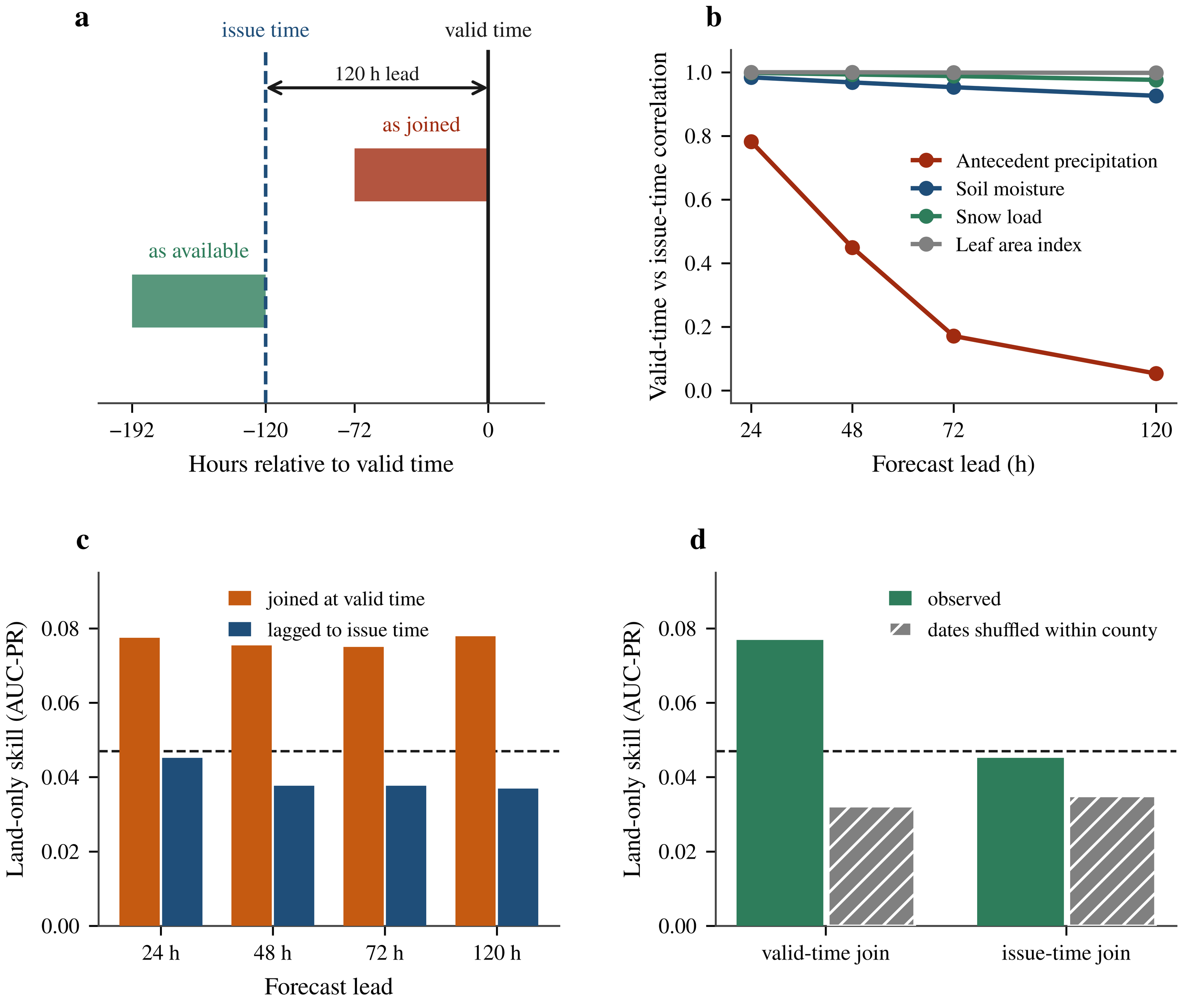}
\caption{\textbf{Temporal leakage in antecedent land surface predictors.} (a) The two feature windows for a forecast issued 120 hours before the valid time. A 72 hour accumulation window ending at the valid time, shown in red, lies entirely after the issue time and therefore contains weather that occurs during the forecast period. The window available to a forecaster, shown in green, ends at the issue time. (b) Correlation between the values of each land surface quantity at valid time and at issue time, by lead time. Antecedent precipitation becomes almost uncorrelated by five days, because its accumulation window, shifted by the lead time, hardly overlaps with itself, whereas slowly varying quantities remain largely preserved. (c) Occurrence skill of the land surface predictors alone under each joining convention. The dashed line is the county climatological reference of 0.047, which the corrected values do not exceed at any lead time. (d) Permutation test, in which the land surface vector of each county is replaced by that of the same county on a randomly chosen different date. This keeps county identity but removes timing information. The permutation removes 58.2\% of the skill of the land surface predictors when they are joined at valid time, and 22.6\% when they are joined at issue time.}
\label{fig:3}
\end{figure}

\subsection{Forecast skill and decision value}

Adding land surface predictors to meteorology improves occurrence skill considerably, from 0.1027 to 0.1207 at 24 h, an increase of 17.5\%, with similar relative gains at 48 and 72 h (Fig. 4a). Based on this result alone, the land surface features appear worth using. The staging decision does not improve in the same way. The change in captured burden from the same addition is +0.028 at 24 h with a 95\% interval of {[}\textminus{}0.003, +0.069{]}, +0.025 at 48 h, +0.011 at 72 h and \textminus{}0.024 at 120 h, and no interval excludes zero (Fig. 4b). An improvement of 17.5\% in ranking skill therefore produces no measurable improvement in the allocation of crews.

Two mechanisms can explain this. First, ranking metrics summarise the whole ordering of counties, whereas a staging rule that selects the highest decile depends only on the top of the ranking. Improvements among counties of lower rank therefore do not reach the decision. Second, skill is estimated over 4.7 million county--windows, whereas captured burden is estimated over 72 high burden windows, so the second estimate is much less precise. Both mechanisms are real, and neither is specific to our setup. In practice, an improvement in skill is therefore not enough evidence that a decision improves. The difference between forecast skill and forecast value is well known in forecast verification (Murphy, 1977; Richardson, 2000), but it is rarely examined in outage prediction, where accuracy improvements are usually reported without assessing their effect on decisions.

Attribution analysis explains this result. Fig. 4c shows the share of mean absolute SHAP attribution (Lundberg and Lee, 2017) for each group of features. Forecast thermodynamic fields contribute 27.3\% and forecast precipitation 16.6\%, while static canopy cover contributes 25.0\%, the second largest share of any group. Prognostic soil and antecedent moisture contribute 9.0\% and snow 6.9\%.

A quarter of the attribution therefore comes from an annual satellite product that describes where trees are, not from conditions that change over time. Because canopy cover hardly changes between years at county scale, it acts as a fixed county characteristic. This explains why the land surface predictors alone do not exceed county climatology (Section 3.2), and why 77\% of their skill at issue time remains after the permutation test (Section 3.2).

\begin{figure}[!htbp]
\centering
\includegraphics[width=\linewidth,height=0.62\textheight,keepaspectratio]{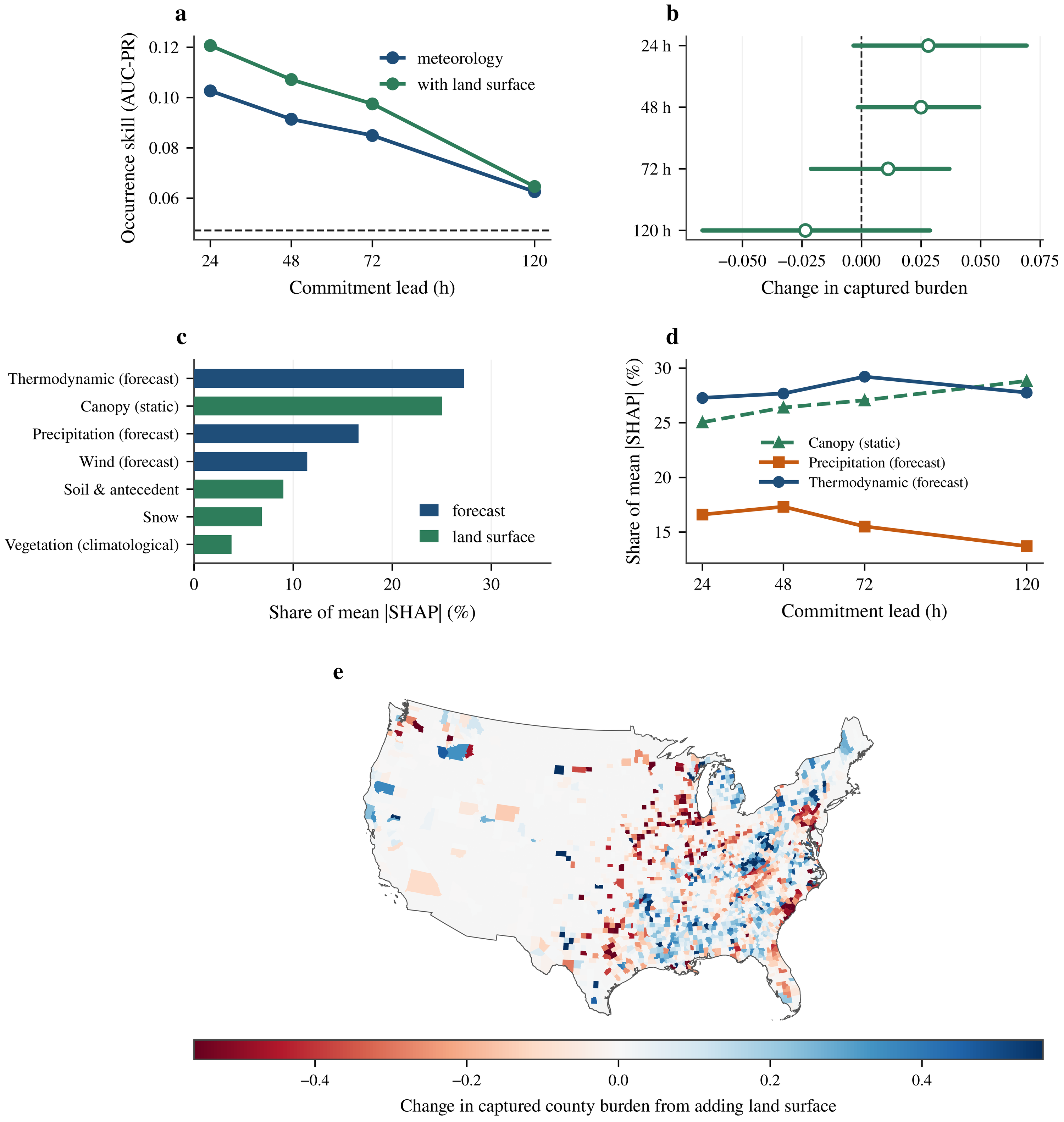}
\caption{\textbf{Forecast skill and decision value.} (a) Occurrence skill against lead time for meteorology alone and with land surface predictors added, evaluated at forecast issue time. The dashed line is the county climatological reference. (b) Change in captured outage burden from adding land surface predictors, with 95\% intervals from a block bootstrap over storm weeks. Open markers denote intervals that include zero. No interval excludes zero at any lead time, so the skill gain in panel (a) does not produce a measurable change in the staging allocation. (c) Share of mean absolute SHAP attribution by feature group at 24 hours lead time, with forecast groups in blue and land surface groups in green. Static canopy cover has the second largest share of any group, although it is an annual product that hardly changes over time. (d) The same decomposition across lead times, for the groups that change. Solid lines are forecast groups and dashed lines are land surface groups. As the forecast becomes less accurate, the model relies more on static canopy and less on forecast precipitation. (e) Change in captured burden in each county from adding land surface predictors, at 24 hours lead time. Land surface predictors improve the allocation in 748 counties and worsen it in 1,056, so the national mean close to zero in panel (b) is a mixture of opposite effects rather than a small effect. Counties in the lowest quartile of observed burden are not shaded, because captured burden is unstable where burden is negligible.}
\label{fig:4}
\end{figure}

Fig. 4d shows how this composition changes with lead time. The canopy share rises from 25.0\% at 24 h to 28.8\% at 120 h, while the share of forecast precipitation falls from 16.6\% to 13.7\%, and the share of forecast wind stays almost constant. As the forecast becomes less accurate, the model relies more on static exposure and less on forecast precipitation, a field that is generally harder to forecast at long lead times.

The national change close to zero hides a mixture of opposite effects rather than a small effect. Adding land surface predictors improves captured burden in 748 counties and worsens it in 1,056, while the remaining 687 change by less than one percentage point (Fig. 4e). The negative national mean at 120 h is consistent with land surface conditions that are five days old adding noise that pushes better ranked counties out of the staged set. The point estimates suggest that an operational system could use land surface predictors only at some lead times, keeping them from 24 to 72 h and removing them at 120 h. However, none of the underlying intervals excludes zero, so this is a direction suggested by the point estimates rather than an established result.

\subsection{Prediction intervals under a change in hazard mix}

Conformalized quantile regression calibrated on 2023 reaches a marginal coverage of 0.861 in 2024, against a nominal 0.90. This small shortfall is consistent with the mild change in distribution expected between consecutive years. The occurrence component is also well calibrated across four orders of magnitude of predicted probability (Fig. 5b).

Coverage by hazard class shows a different picture (Fig. 5a). Tropical cyclone events reach only 0.476, flood events 0.584 and convective events 0.777, while winter and wind events exceed 0.83. A method that appears adequate overall is therefore poorly calibrated for the events that caused the largest outages in the test year, and a utility relying on the marginal guarantee would be overconfident during a hurricane.

Coverage by decile of predicted severity shows a second, smaller failure (Fig. 5c). Coverage declines from 0.900 in the lowest decile to 0.750 in the highest, so intervals are least reliable for the largest predicted events. We stratify by predicted rather than observed severity, because conditioning on the observed outcome would select on the quantity being evaluated.

\begin{figure}[!htbp]
\centering
\includegraphics[width=\linewidth,height=0.62\textheight,keepaspectratio]{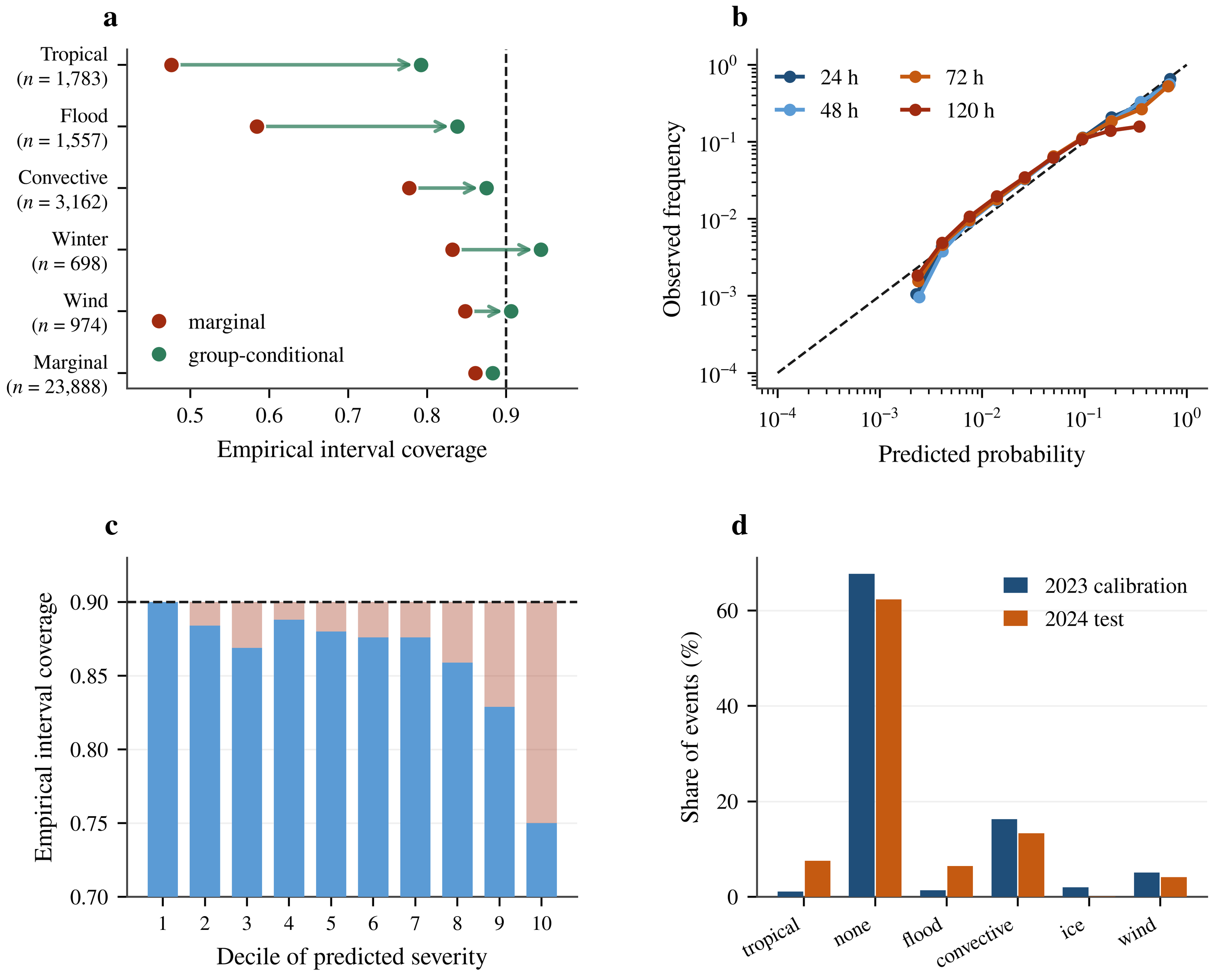}
\caption{\textbf{Predictive interval coverage under hazard mix shift.} (a) Coverage of nominal 90\% intervals by hazard class, with marginal calibration and with calibration by hazard class. Arrows show the direction of the correction, and sample sizes are the numbers of events in the test year. The marginal coverage of 0.861 hides a coverage of only 0.476 for tropical cyclone events. (b) Reliability of the occurrence component, plotted on logarithmic axes because a base rate near 2\% places almost all predicted probabilities in the lowest decile on linear axes. (c) Coverage by decile of predicted severity, with the shortfall from the nominal level shaded above each bar. Stratifying by predicted rather than observed severity avoids selecting on the quantity being evaluated. (d) Composition of event windows by hazard class in the calibration and test years. The calibration year had no major hurricane landfall on the US mainland and the test year had three, which violates the exchangeability condition on which the marginal coverage guarantee depends.}
\label{fig:5}
\end{figure}

The failure is caused by the change in hazard composition between the calibration and test years (Fig. 5d). The 2023 calibration year contained 448 tropical cyclone event windows and no major landfall on the US mainland, whereas the 2024 test year contained 1,783, including Hurricanes Beryl, Helene and Milton. Conformal coverage guarantees require the calibration and test data to be exchangeable, and a year with three major landfalls is not exchangeable with a year with none.

This is a general risk for conformal methods applied to weather impacts. The mix of hazards varies strongly from year to year, so a calibration period of one or two years cannot be assumed to be representative. The failure is not a defect of conformal prediction but a violation of its stated condition.

Calibrating a separate offset within each hazard class, following Equation (7), corrects the failure. Tropical coverage rises from 0.476 to 0.792, flood from 0.584 to 0.838 and convective from 0.777 to 0.875, and marginal coverage improves from 0.861 to 0.883 (Fig. 5a). The cost is an 11\% increase in mean interval width, from 1.967 to 2.179 in units of log severity. Tropical coverage remains below the nominal level after the correction. This is expected, because the correction can only use the 448 tropical calibration events, about a quarter of the 1,783 in the test year.

This correction is a diagnostic rather than an operational method, because hazard class is assigned from storm reports compiled after the event. It shows that the coverage failure is conditional and can be corrected in principle. An operational version would need a hazard class derived from the forecast fields themselves, which is a natural extension but is not attempted here.

Adding the training year to the calibration set is an attractive way to enlarge a small calibration sample, but it reduces marginal coverage from 0.861 to 0.791. Residuals on the training data are smaller than on new data, so adding them reduces the conformal offset from 0.181 to 0.042 and narrows every interval. Keeping the calibration data separate from the training data is therefore essential, and the loss from ignoring this is larger than the change in distribution that the added data were meant to address.

\subsection{Skill by hazard and region}

Skill differs considerably between hazard classes (Table 8). Winter and wind hazards are predicted best under both measures, with lifts of 5.03 and 4.67 and normalised skill of 0.438 and 0.387. Both are driven by quantities that the forecast represents directly and that act directly on power lines and vegetation, and both tend to be organised at large scales, which makes them predictable at county scale. The ranking of the other three hazards depends on the measure. By lift, convective events rank above flood and tropical events, but after normalising for the base rate, tropical events rank above the other two and convective events rank last. This difference is informative. Lift is bounded by the reciprocal of the base rate, so the tropical class, with a base rate of 0.170, cannot exceed a lift of 5.9 however well it is predicted. Its lift of 2.71 therefore understates a normalised skill of 0.351, which ranks third. Convective events show the opposite pattern. Their favourable lift comes from a low base rate, and normalisation reduces it to the lowest value in the table. This is consistent with convective damage occurring at scales much smaller than a county, which a forecast at county resolution cannot locate well. Flood ranks low under both measures, which we attribute to its dependence on river routing and on structural exposure, neither of which the model represents.

\begin{table}[!htbp]
\centering
\small
\caption{\textbf{Skill by hazard class.} Event windows of each hazard are compared with control windows from the same counties and calendar months. Event windows are county--6 h cells at or above the severity threshold in the test year, not the events of Section 2.2. Lift is the ratio of the area under the precision--recall curve to the base rate within this matched pool. Lift is bounded by the reciprocal of the base rate and therefore cannot be compared directly between classes with very different base rates. The final column reports (AUC-PR \textminus{} base) / (1 \textminus{} base), which is not bounded in this way.}
\label{tab:8}
\begin{tabular}{@{}>{\raggedright\arraybackslash}p{\dimexpr 0.1741\linewidth-2\tabcolsep\relax}>{\raggedright\arraybackslash}p{\dimexpr 0.1548\linewidth-2\tabcolsep\relax}>{\raggedright\arraybackslash}p{\dimexpr 0.1838\linewidth-2\tabcolsep\relax}>{\raggedright\arraybackslash}p{\dimexpr 0.1257\linewidth-2\tabcolsep\relax}>{\raggedright\arraybackslash}p{\dimexpr 0.1451\linewidth-2\tabcolsep\relax}>{\raggedright\arraybackslash}p{\dimexpr 0.0870\linewidth-2\tabcolsep\relax}>{\raggedright\arraybackslash}p{\dimexpr 0.1295\linewidth-2\tabcolsep\relax}@{}}
\toprule
Hazard & Pool size & Event Windows & Base rate & AUC-PR & Lift & Normalised \\
\midrule
Winter & 7120 & 698 & 0.098 & 0.493 & 5.03 & 0.438 \\
Wind & 10241 & 974 & 0.095 & 0.445 & 4.67 & 0.387 \\
Convective & 38292 & 3162 & 0.083 & 0.321 & 3.89 & 0.260 \\
Flood & 11292 & 1557 & 0.138 & 0.378 & 2.74 & 0.278 \\
Tropical & 10488 & 1783 & 0.170 & 0.461 & 2.71 & 0.351 \\
\bottomrule
\end{tabular}
\end{table}

Skill in the western United States is much lower than elsewhere, and the gap depends on the hazard. Winter hazard lift in the West is 3.35, compared with 5.26 in the Midwest and 6.46 in the Northeast, and the South has too few matched events for a stable estimate. Wind hazard lift in the West is 4.84, the highest of any region, above 4.45 in the South and 4.13 in the Northeast. A regional failure across all hazards would suggest a general problem with western data or geography. Because western wind performance is the best in the country while western winter performance is the worst, the problem appears to be specific to outages in the cold season.

Three possible explanations were tested, and none is supported. The first was missing snow load predictors. Adding the snow variables described in Section 2.5 raised winter lift in the Northeast from 5.78 to 6.46 and in the Midwest from 4.75 to 5.26, but changed it in the West by only \textminus{}0.05. The second was the quality of storm attribution. In the West, 21\% of events match no storm report, twice the rate elsewhere. However, restricting the evaluation to counties whose events all match reports in the same county gives a western winter lift of 2.25, which is lower rather than higher. The third was county size. Western counties average 5,321 km\textsuperscript{2}, compared with 1,469 to 1,669 km\textsuperscript{2} elsewhere, and aggregation over large areas might dilute a local signal. However, within the Northeast the quartile of largest counties reaches a lift of 7.37, higher than the third quartile, so large counties are not harder to predict in themselves.

We therefore report the western winter gap as unexplained. Possible causes that cannot be tested with these data include ice accretion, which is poorly represented in both the forecast fields and the storm record, and the effect of complex terrain on wind and snow, which a 0.25\textdegree{} forecast cannot resolve. The results reported here therefore apply with most confidence to the eastern, midwestern and southern United States, and to wind hazards in the West.

\subsection{Economic value}

Fig. 6a shows the potential economic value from Equation (10) as a function of the user's cost--loss ratio. Value peaks at a cost--loss ratio of 0.022, close to the climatological event frequency as theory requires, and declines from 0.377 at 24 h to 0.302 at 120 h. A utility can locate itself on this axis from the ratio of its own protection cost to its own loss, and read off whether the forecast is better than climatology and by how much. This measure requires no assumption about the value of lost load, which is why we present it first. The 20\% decline in peak value between one and five days is the same result as in Section 3.1, expressed as economic value rather than as captured burden.

\begin{figure}[!htbp]
\centering
\includegraphics[width=\linewidth,height=0.62\textheight,keepaspectratio]{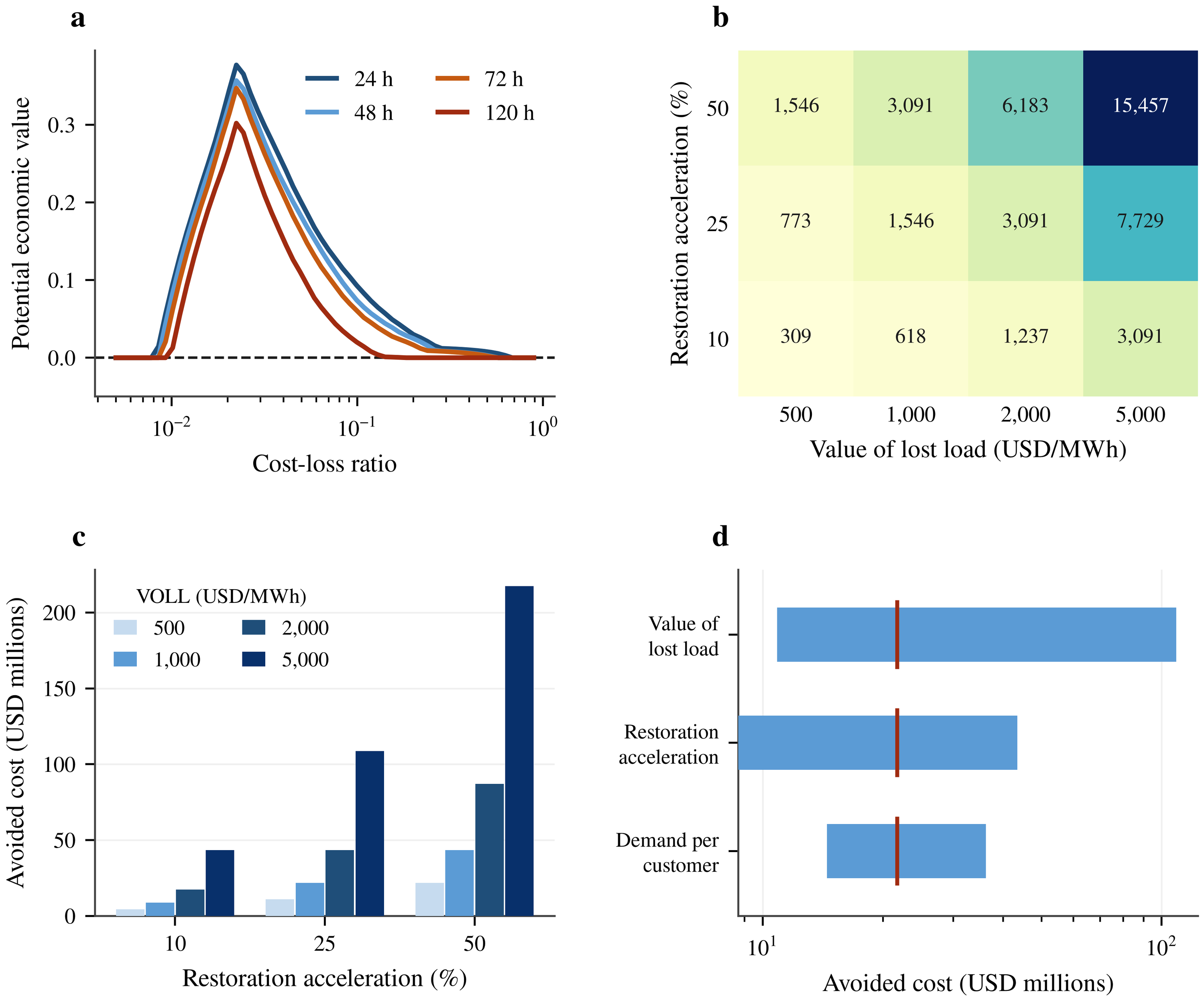}
\caption{\textbf{Economic value of commitment timing.} (a) Potential economic value against the user's cost--loss ratio, by lead time. Value peaks near the climatological event frequency, as theory requires, and declines from 0.377 at 24 hours to 0.302 at 120 hours. This measure requires no assumption about the value of lost load, and a utility can locate itself on the horizontal axis from the ratio of its own protection cost to its own loss. (b) Break-even staging cost per county-day at 72 hours commitment, across values of lost load and restoration acceleration. Staging crews in advance pays when the staging cost per county is below the value shown. A break-even value is reported instead of an assumed rate because published daily rates for storm crews are not available in citable form. (c) Value of committing at 72 rather than 120 hours, by restoration acceleration and value of lost load. (d) Sensitivity of this value to each assumption, ordered by influence. The vertical line marks the central case. The value of lost load changes the estimate by an order of magnitude and restoration acceleration by a factor of five, so the monetary values show the scale of the effect rather than precise estimates.}
\label{fig:6}
\end{figure}

Across the 72 high burden windows of the test year, 598 million customer hours of outage occurred. At a mean demand of 1.2 kW per customer, this equals approximately 718,000 MWh of unserved energy. Between commitment at 72 h and at 120 h, the customer hours falling in staged counties differ by 72.5 million.

Using Equation (11), this difference is worth 21.7 million USD for a restoration acceleration of 25\% and a value of lost load of 1,000 USD/MWh. Across the full range of assumptions it varies from 4.3 to 217 million USD (Fig. 6c). This is the value of the timing decision alone. It is not the quantity that shows whether staging crews in advance pays for itself.

That question requires the avoided cost of staging compared with not staging, which is larger. Staging at 72 h places crews in counties that account for 178.8 million customer hours of observed outage, worth 53.7 million USD under the same assumptions. The corresponding crew commitment is

\begin{equation}
D = m \cdot d \cdot \sum_{t \in W} K_{t}
\label{eq:12}
\end{equation}
\noindent where \(K_{t}\) is the number of counties staged in window \(t\), \(d\) is the number of days crews are held, and \(m\) is the number of crews assigned per staged county. Staging the 10\% of reporting counties ranked highest in each of the 72 high burden windows gives a mean \(K_{t}\) of 161 and a total of 11,570 county stagings. With \(d = 3\) days for a 72 h commitment, this amounts to 34,710 county-days of staging. Dividing the avoided cost by this figure gives a break-even cost of 1,546 USD per county-day under the central assumptions, rising to 15,457 USD for a value of lost load of 5,000 USD/MWh and 50\% acceleration (Fig. 6b). Per crew, the figure is 1,546/m, so a utility that assigns two crews to each staged county has a break-even cost of 773 USD per crew and day. We report the cost per county-day because \(m\) is a staffing choice rather than a property of the forecast. Equation (12) treats the staging in each window as a separate commitment. This overstates \(D\) when several windows belong to the same storm, and therefore understates the break-even cost.

Fig. 6d orders the assumptions by their influence. Across its plausible range, the value of lost load changes the estimate by an order of magnitude, restoration acceleration by a factor of five and demand per customer by a factor of two and a half. The monetary values therefore show the scale of the effect rather than precise estimates, and the analysis should be read mainly through Fig. 6a, which needs none of these assumptions.

Restoration acceleration in particular cannot be measured from our data. Staging crews in advance does not prevent outages. It shortens restoration by an amount that depends on travel distance, type of damage and crew productivity, none of which the outage record contains. We therefore report avoided customer hours under a stated acceleration and do not claim avoided outages.

\subsection{Limitations}

The results have several limitations. First, missing reports in the outage record are not random. Event rates rise from 0.013 to 0.046 across quartiles of reporting sparsity, because utilities that report rarely tend to report mainly during incidents. Predicted probabilities are therefore conditional on reporting, and comparisons between counties are valid only among counties with similar reporting. This does not drive the land surface results, because the gain from land surface predictors decreases across these quartiles, from +0.019 in counties that report most often to +0.010 in those that report least.

Second, severity prediction is weak, with a conditional coefficient of determination of 0.044 on event windows. Occurrence is therefore the main outcome. Because the staging analysis ranks counties by occurrence probability alone, this weakness does not affect the decision results. Under proper scoring rules, the severity component is better than a climatological reference, with a continuous ranked probability skill score of +0.042 and Brier skill scores of +0.046 and +0.055 at severity thresholds of 0.10 and 0.25. However, it is not better than a reference conditioned on hazard class, and its predicted distributions are too wide.

Third, three years of data with a single training year limit how far the results can be generalised. The 2024 test year had an unusually active hurricane season. This is central to the calibration result in Section 3.4, but it also limits how far that result applies to other years. In addition, county and 6 h resolution suit the staging of crews before a storm but not restoration planning at the level of individual feeders. The outage record is reported by county, so finer resolution is not available from this source.

Fourth, the gust proxy uses an assumed shear coefficient of 0.5. Fitting this coefficient against ERA5 instantaneous gusts over the full period gives a pooled value of 1.382, which reduces the root mean square error from 3.15 to 2.34 m/s. Values for individual hazards range from 1.38 for ice storms to 2.21 for wildfire. The proxy as implemented therefore underestimates gust speed, and by different amounts for different hazards. Refitting the coefficient would be a simple improvement that we did not make. The proxy is one of 37 meteorological predictors, and the occurrence results are reported for the model as actually fitted.

Finally, only one pair of forecast models was used, chosen by the variables available in the archive. Repeating the analysis with another AI weather model would show whether the decline in skill reported here is specific to these models or common to AI forecasts in general. For wind, however, the comparison in Section 3.1 suggests that the gain from a different forecast model would be small compared with the gain from a gust field, which the archive does not provide.

\section{Conclusions}

This study used operational AI weather forecasts to predict outages at county scale across the contiguous United States. The forecasts were evaluated by the crew commitment decisions they support rather than by prediction accuracy alone. The results show that committing crews five days before a storm captures significantly less outage burden than committing three days before. Captured outage burden decreases steadily with lead time, and the loss between three and five days is significant in every analysis. Within the first three days, the loss is significant when burden is limited to outages that have not yet started, but not when outages already in progress are included. The apparent flexibility in this period is therefore partly caused by event duration rather than by the forecast. Committing early with all requested crews outperforms committing late with fewer crews, and we report the crew availability ratio at which the two options break even.

Antecedent land surface predictors must be evaluated at the forecast issue time. When their accumulation window ends at the valid time, they include weather from the forecast period, which inflates their skill enough to reverse the conclusion that the uncorrected data appear to support. The problem is concentrated in quantities computed over rolling windows, while slowly varying quantities are largely unaffected. The correlation between values at valid time and at issue time identifies the affected features, and this simple diagnostic can be applied to any impact forecasting study that uses antecedent conditions. Gains in forecast skill do not necessarily improve the decisions that forecasts support. Land surface predictors improved occurrence skill considerably but did not significantly change captured burden. Attribution analysis linked this gap to the static component of these predictors, whose share of attribution increases as the weather forecast becomes less accurate.

Conformal prediction intervals need calibration by hazard class when the mix of hazards changes between years. Intervals with correct overall coverage failed for tropical cyclones, because the calibration year had no major hurricane landfall on the US mainland and the test year had three. Calibration by hazard class corrected coverage at a small cost in interval width. Because hazard classes are assigned after the event, this correction is a diagnostic rather than an operational method.

Overall, the value of forecasts for crew commitment can be reported without assuming a value of lost load, using potential economic value across a range of cost--loss ratios and the break-even staging cost per county-day. Future work should address three points. The first concerns the variables in the archive rather than the forecast models. Among specifications that use wind only, replacing forecast wind with analysis wind improves occurrence skill by about 15\%, while adding an instantaneous gust field improves it by about 33\%. This suggests that archiving gusts may matter more than improving the fields already archived. The comparison has limits, because both specifications are close to county climatology and an archived gust forecast would itself become less accurate with lead time. Second, the correction by hazard class needs a hazard class derived from the forecast before it can be used operationally. Third, the lower winter skill in the western United States, which remained unexplained after three tests, points to processes that neither the forecast fields nor the storm record represent well, and will require additional data.

\subsection*{Data and Code Availability}

All source data are publicly available: EAGLE-I outage records from Oak Ridge National Laboratory (Brelsford et al., 2024), the NOAA/CIRA AIWP forecast archive (Radford et al., 2025), ERA5-Land reanalysis from the Copernicus Climate Data Store (Muñoz-Sabater et al., 2021), NLCD tree canopy cover (Dewitz, 2024), the NOAA Storm Events Database, and American Community Survey housing-unit estimates (U.S. Census Bureau, 2023). The code used to build the county--window dataset, fit the models and produce every table and figure, together with the feature manifest and the processed county--window tables, is archived at Zenodo (https://doi.org/10.5281/zenodo.23326163).

\subsection*{Declaration of competing interest}

The authors declare that they have no competing interests for this study.

\section*{References}
\begin{list}{}{\setlength{\leftmargin}{1.5em}\setlength{\itemindent}{-1.5em}\setlength{\itemsep}{2pt}\setlength{\parsep}{0pt}}
\small
\item Andresen, A.X., Kurtz, L.C., Hondula, D.M., Meerow, S., Gall, M., 2023. Understanding the social impacts of power outages in North America: a systematic review. Environ. Res. Lett. 18, 053004. \url{https://doi.org/10.1088/1748-9326/acc7b9}
\item Arab, A., Khodaei, A., Khator, S.K., Ding, K., Emesih, V.A., Han, Z., 2015. Stochastic Pre-hurricane Restoration Planning for Electric Power Systems Infrastructure. IEEE Transactions on Smart Grid 6, 1046--1054. \url{https://doi.org/10.1109/TSG.2015.2388736}
\item Arora, P., Ceferino, L., 2023. Probabilistic and machine learning methods for uncertainty quantification in power outage prediction due to extreme events. Natural Hazards and Earth System Sciences 23, 1665--1683. \url{https://doi.org/10.5194/nhess-23-1665-2023}
\item Arora, P., Ceferino, L., Dobler, G., 2026. Satellite based inference for disaster induced power outages using daily nighttime light observations. npj Nat. Hazards. \url{https://doi.org/10.1038/s44304-026-00270-z}
\item Azizi, M., Zhang, X., Yasenpoor, T., Cerrai, D., Song, D., Anagnostou, E., 2026. Addressing the data imbalance issue in machine learning modeling of rare and disruptive outage events. Sci Rep 16, 8876. \url{https://doi.org/10.1038/s41598-026-41838-z}
\item Bi, K., Xie, L., Zhang, H., Chen, X., Gu, X., Tian, Q., 2023. Accurate medium-range global weather forecasting with 3D neural networks. Nature 619, 533--538. \url{https://doi.org/10.1038/s41586-023-06185-3}
\item Bodnar, C., Bruinsma, W.P., Lucic, A., Stanley, M., Allen, A., Brandstetter, J., Garvan, P., Riechert, M., Weyn, J.A., Dong, H., Gupta, J.K., Thambiratnam, K., Archibald, A.T., Wu, C.-C., Heider, E., Welling, M., Turner, R.E., Perdikaris, P., 2025. A foundation model for the Earth system. Nature 641, 1180--1187. \url{https://doi.org/10.1038/s41586-025-09005-y}
\item Bonev, B., Kurth, T., Hundt, C., Pathak, J., Baust, M., Kashinath, K., Anandkumar, A., 2023. Spherical Fourier neural operators: learning stable dynamics on the sphere, in: Proceedings of the 40th International Conference on Machine Learning, ICML'23. JMLR.org, Honolulu, Hawaii, USA, pp. 2806--2823.
\item Brelsford, C., Tennille, S., Myers, A., Chinthavali, S., Tansakul, V., Denman, M., Coletti, M., Grant, J., Lee, S., Allen, K., Johnson, E., Huihui, J., Hamaker, A., Newby, S., Medlen, K., Maguire, D., Dunivan Stahl, C., Moehl, J., Redmon, D., Sanyal, J., Bhaduri, B., 2024. A dataset of recorded electricity outages by United States county 2014--2022. Sci Data 11, 271. \url{https://doi.org/10.1038/s41597-024-03095-5}
\item Cerrai, D., Wanik, D.W., Bhuiyan, M.A.E., Zhang, X., Yang, J., Frediani, M.E.B., Anagnostou, E.N., 2019. Predicting Storm Outages Through New Representations of Weather and Vegetation. IEEE Access 7, 29639--29654. \url{https://doi.org/10.1109/ACCESS.2019.2902558}
\item Charlton-Perez, A.J., Dacre, H.F., Driscoll, S., Gray, S.L., Harvey, B., Harvey, N.J., Hunt, K.M.R., Lee, R.W., Swaminathan, R., Vandaele, R., Volonté, A., 2024. Do AI models produce better weather forecasts than physics-based models? A quantitative evaluation case study of Storm Ciarán. npj Clim Atmos Sci 7, 93. \url{https://doi.org/10.1038/s41612-024-00638-w}
\item DeMaria, M., Franklin, J.L., Chirokova, G., Radford, J., DeMaria, R., Musgrave, K.D., Ebert-Uphoff, I., 2025. An Operations-Based Evaluation of Tropical Cyclone Track and Intensity Forecasts from Artificial Intelligence Weather Prediction Models. Artificial Intelligence for the Earth Systems 4. \url{https://doi.org/10.1175/AIES-D-24-0085.1}
\item Dewitz, J. (2024). National Land Cover Database (NLCD) tree canopy cover products {[}dataset{]}. U.S. Geological Survey data release. \url{https://doi.org/10.5066/P9JZ7AO3}
\item Do, V., McBrien, H., Flores, N.M., Northrop, A.J., Schlegelmilch, J., Kiang, M.V., Casey, J.A., 2023. Spatiotemporal distribution of power outages with climate events and social vulnerability in the USA. Nat Commun 14, 2470. \url{https://doi.org/10.1038/s41467-023-38084-6}
\item Esmaeilian, F., Zhang, X., Azizpourshoubi, F., Astitha, M., Anagnostou, E., 2026. Assessing the Propagation of Weather Forecast Errors into Power Outage Predictions. Forecasting 8. \url{https://doi.org/10.3390/forecast8040062}
\item Fatehi, N., Biswas, A.K., Nazari, M.H. (2025). Predictive modeling of power outages during extreme events: integrating weather and socio-economic factors. arXiv preprint arXiv:2512.22699. \url{https://doi.org/10.48550/arXiv.2512.22699}
\item Flores, N.M., Northrop, A.J., Do, V., Gordon, M., Jiang, Y., Rudolph, K.E., Hernández, D., Casey, J.A., 2024. Powerless in the storm: Severe weather-driven power outages in New York State, 2017--2020. PLOS Climate 3, e0000364. \url{https://doi.org/10.1371/journal.pclm.0000364}
\item Frakes, E., Wu, Y., French, R.H., Li, M., 2026. GeoOutageKG: A Multimodal Geospatiotemporal Knowledge Graph for~Multiresolution Power Outage Analysis, in: Garijo, D., Kirrane, S., Salatino, A., Shimizu, C., Acosta, M., Nuzzolese, A.G., Ferrada, S., Soulard, T., Kozaki, K., Takeda, H., Gentile, A.L. (Eds.), The Semantic Web -- ISWC 2025. Springer Nature Switzerland, Cham, pp. 221--239. \url{https://doi.org/10.1007/978-3-032-09530-5\_13}
\item Ganz, S.C., Duan, C., Ji, C., 2023. Socioeconomic vulnerability and differential impact of severe weather-induced power outages. PNAS Nexus 2, pgad295. \url{https://doi.org/10.1093/pnasnexus/pgad295}
\item Grinsztajn, L., Oyallon, E., Varoquaux, G., 2022. Why do tree-based models still outperform deep learning on typical tabular data?, in: Advances in Neural Information Processing Systems 35, Datasets and Benchmarks Track. Curran Associates, Inc., pp. 507--520.
\item Guikema, S.D., Nateghi, R., Quiring, S.M., Staid, A., Reilly, A.C., Gao, M., 2014. Predicting Hurricane Power Outages to Support Storm Response Planning. IEEE Access 2, 1364--1373. \url{https://doi.org/10.1109/ACCESS.2014.2365716}
\item Guikema, S.D., Quiring, S.M., 2012. Hybrid data mining-regression for infrastructure risk assessment based on zero-inflated data. Reliability Engineering \& System Safety 99, 178--182. \url{https://doi.org/10.1016/j.ress.2011.10.012}
\item Guikema, S.D., Quiring, S.M., Han, S.-R., 2010. Prestorm Estimation of Hurricane Damage to Electric Power Distribution Systems. Risk Analysis 30, 1744--1752. \url{https://doi.org/10.1111/j.1539-6924.2010.01510.x}
\item Han, S.-R., Guikema, S.D., Quiring, S.M., 2009. Improving the Predictive Accuracy of Hurricane Power Outage Forecasts Using Generalized Additive Models. Risk Analysis 29, 1443--1453. \url{https://doi.org/10.1111/j.1539-6924.2009.01280.x}
\item He, J., Wanik, D.W., Hartman, B.M., Anagnostou, E.N., Astitha, M., Frediani, M.E.B., 2017. Nonparametric Tree-Based Predictive Modeling of Storm Outages on an Electric Distribution Network. Risk Analysis 37, 441--458. \url{https://doi.org/10.1111/risa.12652}
\item Hersbach, H., Bell, B., Berrisford, P., Hirahara, S., Horányi, A., Muñoz-Sabater, J., Nicolas, J., Peubey, C., Radu, R., Schepers, D., Simmons, A., Soci, C., Abdalla, S., Abellan, X., Balsamo, G., Bechtold, P., Biavati, G., Bidlot, J., Bonavita, M., De Chiara, G., Dahlgren, P., Dee, D., Diamantakis, M., Dragani, R., Flemming, J., Forbes, R., Fuentes, M., Geer, A., Haimberger, L., Healy, S., Hogan, R.J., Hólm, E., Janisková, M., Keeley, S., Laloyaux, P., Lopez, P., Lupu, C., Radnoti, G., de Rosnay, P., Rozum, I., Vamborg, F., Villaume, S., Thépaut, J.-N., 2020. The ERA5 global reanalysis. Quarterly Journal of the Royal Meteorological Society 146, 1999--2049. \url{https://doi.org/10.1002/qj.3803}
\item Hughes, W., Zhang, W., Cerrai, D., Bagtzoglou, A., Wanik, D., Anagnostou, E., 2022. A Hybrid Physics-Based and Data-Driven Model for Power Distribution System Infrastructure Hardening and Outage Simulation. Reliability Engineering \& System Safety 225, 108628. \url{https://doi.org/10.1016/j.ress.2022.108628}
\item Kapoor, S., Narayanan, A., 2023. Leakage and the reproducibility crisis in machine-learning-based science. Patterns 4, 100804. \url{https://doi.org/10.1016/j.patter.2023.100804}
\item Ke, G., Meng, Q., Finley, T., Wang, T., Chen, W., Ma, W., Ye, Q., Liu, T.-Y., 2017. LightGBM: a highly efficient gradient boosting decision tree, in: Proceedings of the 31st International Conference on Neural Information Processing Systems, NIPS'17. Curran Associates Inc., Red Hook, NY, USA, pp. 3149--3157.
\item Lam, R., Sanchez-Gonzalez, A., Willson, M., Wirnsberger, P., Fortunato, M., Alet, F., Ravuri, S., Ewalds, T., Eaton-Rosen, Z., Hu, W., Merose, A., Hoyer, S., Holland, G., Vinyals, O., Stott, J., Pritzel, A., Mohamed, S., Battaglia, P., 2023. Learning skillful medium-range global weather forecasting. Science 382, 1416--1421. \url{https://doi.org/10.1126/science.adi2336}
\item Lei, S., Chen, C., Li, Y., Hou, Y., 2019. Resilient Disaster Recovery Logistics of Distribution Systems: Co-Optimize Service Restoration With Repair Crew and Mobile Power Source Dispatch. IEEE Transactions on Smart Grid 10, 6187--6202. \url{https://doi.org/10.1109/TSG.2019.2899353}
\item Lundberg, S.M., Lee, S.-I., 2017. A unified approach to interpreting model predictions, in: Proceedings of the 31st International Conference on Neural Information Processing Systems, NIPS'17. Curran Associates Inc., Red Hook, NY, USA, pp. 4768--4777.
\item Ma, J., Li, B., Omitaomu, O.A., Mostafavi, A., 2025. Establishing nationwide power system vulnerability index across US counties using interpretable machine learning. Applied Energy 397, 126360. \url{https://doi.org/10.1016/j.apenergy.2025.126360}
\item Manning, C., Wilkinson, S., Fowler, H.J., Kendon, E.J., 2025. Antecedent rainfall, wind direction and seasonal effects may amplify the risk of wind-driven power outages in the UK. Commun Earth Environ 6, 217. \url{https://doi.org/10.1038/s43247-025-02176-6}
\item McRoberts, D.B., Quiring, S.M., Guikema, S.D., 2018. Improving Hurricane Power Outage Prediction Models Through the Inclusion of Local Environmental Factors. Risk Analysis 38, 2722--2737. \url{https://doi.org/10.1111/risa.12728}
\item Muñoz-Sabater, J., Dutra, E., Agustí-Panareda, A., Albergel, C., Arduini, G., Balsamo, G., Boussetta, S., Choulga, M., Harrigan, S., Hersbach, H., Martens, B., Miralles, D.G., Piles, M., Rodríguez-Fernández, N.J., Zsoter, E., Buontempo, C., Thépaut, J.-N., 2021. ERA5-Land: a state-of-the-art global reanalysis dataset for land applications. Earth System Science Data 13, 4349--4383. \url{https://doi.org/10.5194/essd-13-4349-2021}
\item Murphy, A.H., 1977. The Value of Climatological, Categorical and Probabilistic Forecasts in the Cost-Loss Ratio Situation. Monthly Weather Review 105, 803--816. \url{https://doi.org/10.1175/1520-0493(1977)105\%3C0803:TVOCCA\%3E2.0.CO;2}
\item Nateghi, R., Guikema, S., Quiring, S.M., 2014. Power Outage Estimation for Tropical Cyclones: Improved Accuracy with Simpler Models. Risk Analysis 34, 1069--1078. \url{https://doi.org/10.1111/risa.12131}
\item NOAA National Centers for Environmental Information (2025). Storm Events Database {[}dataset{]}. NOAA National Centers for Environmental Information, Asheville, NC. Dataset identifier gov.noaa.ncdc:C00510. \url{https://www.ncdc.noaa.gov/stormevents/} (accessed {[}1 May 2026{]}).
\item Panteli, M., Mancarella, P., 2015. Influence of extreme weather and climate change on the resilience of power systems: Impacts and possible mitigation strategies. Electric Power Systems Research 127, 259--270. \url{https://doi.org/10.1016/j.epsr.2015.06.012}
\item Pathak, J., Subramanian, S., Harrington, P., Raja, S., Chattopadhyay, A., Mardani, M., Kurth, T., Hall, D., Li, Z., Azizzadenesheli, K., Hassanzadeh, P., Kashinath, K., Anandkumar, A. (2022). FourCastNet: a global data-driven high-resolution weather model using adaptive Fourier neural operators. arXiv preprint arXiv:2202.11214. \url{https://doi.org/10.48550/arXiv.2202.11214}
\item Price, I., Sanchez-Gonzalez, A., Alet, F., Andersson, T.R., El-Kadi, A., Masters, D., Ewalds, T., Stott, J., Mohamed, S., Battaglia, P., Lam, R., Willson, M., 2025. Probabilistic weather forecasting with machine learning. Nature 637, 84--90. \url{https://doi.org/10.1038/s41586-024-08252-9}
\item Ptak, T., Brooks, J., Stock, R., 2025. A systematic review and typology of power outage literature: Critical infrastructure, climate change and social impacts. Renewable and Sustainable Energy Reviews 218, 115778. \url{https://doi.org/10.1016/j.rser.2025.115778}
\item Quiring, S.M., Zhu, L., Guikema, S.D., 2011. Importance of soil and elevation characteristics for modeling hurricane-induced power outages. Nat Hazards 58, 365--390. \url{https://doi.org/10.1007/s11069-010-9672-9}
\item Radford, J.T., Ebert-Uphoff, I., Stewart, J.Q., Musgrave, K.D., DeMaria, R., Tourville, N., Hilburn, K., 2025. Accelerating Community-Wide Evaluation of AI Models for Global Weather Prediction by Facilitating Access to Model Output. Bulletin of the American Meteorological Society 106, E68--E76. \url{https://doi.org/10.1175/BAMS-D-24-0057.1}
\item Regnier, E., 2008. Public Evacuation Decisions and Hurricane Track Uncertainty. Management Science 54, 16--28. \url{https://doi.org/10.1287/mnsc.1070.0764}
\item Regnier, E., Harr, P.A., 2006. A Dynamic Decision Model Applied to Hurricane Landfall. Weather and Forecasting 21, 764--780. \url{https://doi.org/10.1175/WAF958.1}
\item Richardson, D.S., 2001. Measures of skill and value of ensemble prediction systems, their interrelationship and the effect of ensemble size. Quarterly Journal of the Royal Meteorological Society 127, 2473--2489. \url{https://doi.org/10.1002/qj.49712757715}
\item Richardson, D.S., 2000. Skill and relative economic value of the ECMWF ensemble prediction system. Quarterly Journal of the Royal Meteorological Society 126, 649--667. \url{https://doi.org/10.1002/qj.49712656313}
\item Romano, Y., Patterson, E., Candès, E.J., 2019. Conformalized quantile regression, in: Proceedings of the 33rd International Conference on Neural Information Processing Systems. Curran Associates Inc., Red Hook, NY, USA, pp. 3543--3553.
\item Shashaani, S., Guikema, S.D., Zhai, C., Pino, J.V., Quiring, S.M., 2018. Multi-Stage Prediction for Zero-Inflated Hurricane Induced Power Outages. IEEE Access 6, 62432--62449. \url{https://doi.org/10.1109/ACCESS.2018.2877078}
\item Shen, X., Pan, Z., Cerrai, D., Zhang, X., Colorio, C., Anagnostou, E., Song, D. (2026). Empowering power outage prediction with spatially aware hybrid graph neural networks and contrastive learning. arXiv preprint arXiv:2604.04916. \url{https://doi.org/10.48550/arXiv.2604.04916}
\item Stürmer, J., Plietzsch, A., Vogt, T., Hellmann, F., Kurths, J., Otto, C., Frieler, K., Anvari, M., 2024. Increasing the resilience of the Texas power grid against extreme storms by hardening critical lines. Nat Energy 9, 526--535. \url{https://doi.org/10.1038/s41560-023-01434-1}
\item Tansakul, V., Myers, A., Tennille, S., et al. (2023). EAGLE-I power outage data information {[}dataset{]}. Oak Ridge National Laboratory. \url{https://doi.org/10.13139/ORNLNCCS/1975203}
\item Tervo, R., Láng, I., Jung, A., Mäkelä, A., 2021. Predicting power outages caused by extratropical storms. Natural Hazards and Earth System Sciences 21, 607--627. \url{https://doi.org/10.5194/nhess-21-607-2021}
\item U.S. Census Bureau (2023). American Community Survey 5-year estimates, 2018--2022, Table B25001: Housing units {[}dataset{]}. U.S. Department of Commerce. \url{https://data.census.gov} (accessed {[}5 May 2026{]}).
\item Vovk, V., 2013. Conditional validity of inductive conformal predictors. Mach Learn 92, 349--376. \url{https://doi.org/10.1007/s10994-013-5355-6}
\item Wanik, D.W., Anagnostou, E.N., Hartman, B.M., Frediani, M.E.B., Astitha, M., 2015. Storm outage modeling for an electric distribution network in Northeastern USA. Nat Hazards 79, 1359--1384. \url{https://doi.org/10.1007/s11069-015-1908-2}
\item Watson, P.L., Spaulding, A., Koukoula, M., Anagnostou, E., 2022. Improved quantitative prediction of power outages caused by extreme weather events. Weather and Climate Extremes 37, 100487. \url{https://doi.org/10.1016/j.wace.2022.100487}
\item Wilks, D.S., 2001. A skill score based on economic value for probability forecasts. Meteorological Applications 8, 209--219. \url{https://doi.org/10.1017/S1350482701002092}
\item Yang, F., Cerrai, D., Anagnostou, E.N., 2021. The Effect of Lead-Time Weather Forecast Uncertainty on Outage Prediction Modeling. Forecasting 3, 501--516. \url{https://doi.org/10.3390/forecast3030031}
\item Yang, F., Wanik, D.W., Cerrai, D., Bhuiyan, M.A.E., Anagnostou, E.N., 2020. Quantifying Uncertainty in Machine Learning-Based Power Outage Prediction Model Training: A Tool for Sustainable Storm Restoration. Sustainability 12. \url{https://doi.org/10.3390/su12041525}
\item Zhai, C., Chen, T.Y., White, A.G., Guikema, S.D., 2021. Power outage prediction for natural hazards using synthetic power distribution systems. Reliability Engineering \& System Safety 208, 107348. \url{https://doi.org/10.1016/j.ress.2020.107348}
\end{list}

\clearpage
\section*{Supplementary Material}
\addcontentsline{toc}{section}{Supplementary Material}
\setcounter{table}{0}\renewcommand{\thetable}{S\arabic{table}}

\subsection*{Text S1. Treatment of county--windows with no records}

A county--window with no outage records can be treated in three ways. The last observation can be carried forward, the window can be set to zero customers out, or the window can be masked and excluded from the model. Table S1 compares the three options on the full grid of 3,031 counties before coverage screening. Setting missing windows to zero leaves the number of cells at or above the event threshold unchanged at 132,919, but assigns a value of zero to 7.1 million unobserved cells. This assumes an absence of outages that the record does not support. Carrying the last observation forward raises the number of cells above the threshold to 182,313 by extending outages across gaps when no storm was active. The additional events it creates match an independent storm record at about half the rate of the events kept by masking, so most of them are artefacts of missing reports. We therefore adopted masking.

\begin{table}[!htbp]
\centering
\small
\caption{Treatment of county--windows with no reported records, on the full grid of 3,031 counties before coverage screening. Masking excludes the cell from the model, setting to zero treats missing reports as zero customers out, and carrying forward repeats the last observation. Masking is adopted.}
\label{tab:S1}
\begin{tabular}{@{}>{\raggedright\arraybackslash}p{\dimexpr 0.1222\linewidth-2\tabcolsep\relax}>{\raggedright\arraybackslash}p{\dimexpr 0.1566\linewidth-2\tabcolsep\relax}>{\raggedright\arraybackslash}p{\dimexpr 0.1839\linewidth-2\tabcolsep\relax}>{\raggedright\arraybackslash}p{\dimexpr 0.1240\linewidth-2\tabcolsep\relax}>{\raggedright\arraybackslash}p{\dimexpr 0.1634\linewidth-2\tabcolsep\relax}>{\raggedright\arraybackslash}p{\dimexpr 0.2499\linewidth-2\tabcolsep\relax}@{}}
\toprule
Rule & Cells & Unobserved & Zero & Sub-threshold & At or above threshold \\
\midrule
Mask & 13,199,576 & 7,068,665 & 558,876 & 5,439,116 & 132,919 \\
Set to zero & 13,199,576 & 0 & 7,627,541 & 5,439,116 & 132,919 \\
Carry forward & 13,199,576 & 0 & 4,373,065 & 8,644,198 & 182,313 \\
\bottomrule
\end{tabular}
\end{table}

\begin{table}[!htbp]
\centering
\small
\caption{Sensitivity of the event definition to the severity threshold. The minimum duration is 24 h and the gap tolerance is 12 h throughout. The median duration is insensitive to the threshold up to 0.05. The adopted threshold of 0.05 is the lowest at which the detected events are not dominated by small background outages.}
\label{tab:S2}
\begin{tabular}{@{}>{\raggedright\arraybackslash}p{\dimexpr 0.1442\linewidth-2\tabcolsep\relax}>{\raggedright\arraybackslash}p{\dimexpr 0.1346\linewidth-2\tabcolsep\relax}>{\raggedright\arraybackslash}p{\dimexpr 0.2596\linewidth-2\tabcolsep\relax}>{\raggedright\arraybackslash}p{\dimexpr 0.2404\linewidth-2\tabcolsep\relax}>{\raggedright\arraybackslash}p{\dimexpr 0.2212\linewidth-2\tabcolsep\relax}@{}}
\toprule
Threshold & Events & Counties with \(\geq\)1 event & Median duration (h) & Median peak severity \\
\midrule
0.005 & 32,985 & 2,439 & 30 & 0.061 \\
0.01 & 18,926 & 2,361 & 30 & 0.097 \\
0.02 & 10,758 & 2,189 & 30 & 0.150 \\
0.05 & 5,004 & 1,776 & 30 & 0.266 \\
0.1 & 2,679 & 1,360 & 36 & 0.388 \\
\bottomrule
\end{tabular}
\end{table}

\subsection*{Text S2. Variation of land surface features between years}

For each land surface feature \(j\), we divided its standard deviation across years, at a fixed county \(c\) and day of year \(d\), by its overall standard deviation,

\begin{equation}
\nu_{j} = \frac{\mathbb{E}_{c,d}\left[ \mathrm{sd}_{year}\, l_{j}(c, d, \cdot) \right]}{\mathrm{sd}\left[ l_{j} \right]}
\tag{S1}\label{eq:S1}
\end{equation}

A field prescribed from a repeating climatology gives \(\nu_{j} \approx 0\), whereas a field that responds to actual conditions does not. The canopy descriptors give \(\nu_{j}\) between 0.004 and 0.029, as expected for an annual product. Leaf area index gives \(\nu_{j}\)= 0.005 for high vegetation and 0.002 for low vegetation, because ERA5 prescribes leaf area index from a monthly climatology instead of assimilating satellite retrievals. These features therefore represent the seasonal cycle and not the actual state of vegetation in a given season. Soil moisture, precipitation and snow give \(\nu_{j}\) between 0.078 and 0.616 and change with actual conditions.

\subsection*{Text S3. Comparison of model classes}

We compared the gradient boosting tree model with a regularised linear model and a multilayer perceptron. The tree model was fitted with the 66 features of the main analysis and with added lagged features, features from adjacent counties, or both. The multilayer perceptron was fitted with the 66 features and with each addition separately, and the linear model with the 66 features only. All models were evaluated at 24 h lead on the 2024 test year (Table S3). The neural network performed worse than the tree model with the same features, at 0.102 against 0.123. For this dataset and setup the tree model is therefore the better choice, although the comparison does not establish whether another architecture could extract more from the same predictors. Adding lagged features raised the skill of the tree model by 5.4\% to 0.130, and adding features from adjacent counties raised it by 1.5\%. The neural network did not benefit from either addition, and its severity component performed worse than predicting the mean.

\begin{table}[!htbp]
\centering
\small
\caption{Comparison of model classes at 24 h lead on the 2024 test year. AUC-PR is the area under the precision--recall curve for occurrence, and severity R\textsuperscript{2} is conditional on occurrence. The tree models in this comparison use 1,000 estimators with early stopping after 80 rounds. The comparison joins land surface predictors at valid time, so its reference AUC-PR (0.1233) is higher than the issue-time value in the main text (0.1207 at 24 h); all models receive the same inputs, so the comparison between model classes is unaffected.}
\label{tab:S3}
\begin{tabular}{@{}>{\raggedright\arraybackslash}p{\dimexpr 0.3462\linewidth-2\tabcolsep\relax}>{\raggedright\arraybackslash}p{\dimexpr 0.1538\linewidth-2\tabcolsep\relax}>{\raggedright\arraybackslash}p{\dimexpr 0.1827\linewidth-2\tabcolsep\relax}>{\raggedright\arraybackslash}p{\dimexpr 0.1827\linewidth-2\tabcolsep\relax}>{\raggedright\arraybackslash}p{\dimexpr 0.1346\linewidth-2\tabcolsep\relax}@{}}
\toprule
Model & Features & AUC-PR & Brier score & Severity R\textsuperscript{2} \\
\midrule
Gradient boosting (reference) & 66 & 0.1233 & 0.02108 & 0.031 \\
Gradient boosting + lagged & 77 & 0.1299 & 0.02097 & 0.033 \\
Gradient boosting + adjacent & 71 & 0.1251 & 0.02107 & 0.032 \\
Gradient boosting + both & 82 & 0.1274 & 0.02100 & 0.032 \\
Regularised linear & 66 & 0.1124 & 0.02128 & 0.022 \\
Multilayer perceptron & 66 & 0.1017 & 0.02146 & \textminus{}0.072 \\
Multilayer perceptron + lagged & 77 & 0.0992 & 0.02166 & \textminus{}0.104 \\
Multilayer perceptron + adjacent & 71 & 0.1022 & 0.02150 & \textminus{}0.090 \\
\bottomrule
\end{tabular}
\end{table}

\end{document}